\documentclass[MSNbibl,nameyear,dvips]{arxstspdf}
\usepackage{flushend}
\usepackage{stfloats}
\usepackage{graphicx,url,breakurl}

\volume{29}
\issue{3}
\pubyear{2014}
\firstpage{439}
\lastpage{457}
\doi{10.1214/14-STS489} 
\docsubty{FLA}



\begin{document}
\begin{frontmatter}

\title{A Conversation with Donald B. Rubin}
\runtitle{A Conversation with D. B. Rubin}

\begin{aug}
\author[a]{\fnms{Fan}~\snm{Li}\ead[label=e1]{fli@stat.duke.edu}}
\and
\author[b]{\fnms{Fabrizia}~\snm{Mealli}\corref{}\ead[label=e2]{mealli@disia.unifi.it}}
\runauthor{F. Li and F. Mealli}

\affiliation{Duke University and University of Florence}

\address[a]{Fan Li is Assistant Professor,
Department of Statistical Science, Duke University, Durham, North Carolina 27708-0251, USA \printead{e1}.}
\address[b]{Fabrizia Mealli is Professor, Department of
Statistics, Computer Science, Applications, University of Florence, Viale Morgagni 59, Florence 50134, Italy \printead{e2}.}
\end{aug}

\begin{abstract}
Donald Bruce Rubin is John L. Loeb Professor of
Statistics at Harvard University. He has made fundamental contributions to
statistical methods for missing data, causal inference, survey sampling,
Bayesian inference, computing and applications to a wide range of
disciplines, including psychology, education, policy, law, economics,
epidemiology, public health and other social and biomedical sciences.
\end{abstract}

\end{frontmatter}

Don was born in Washington, D.C. on December 22, 1943, to Harriet and Allan
Rubin. One year later, his family moved to Evanston, Illinois, where he
grew up. He developed a keen interest in physics and mathematics in high
school. In 1961, he went to college at Princeton University, intending to
major in physics, but graduated in psychology in 1965. He began graduate
school in psychology at Harvard, then switched to Computer Science (MS,
1966) and eventually earned a Ph.D. in Statistics under the direction of Bill
Cochran in 1970. After graduating from Harvard, he taught for a year in
Harvard's Department of Statistics, and then in 1971 he began working at
Educational Testing Service (ETS) and served as a visiting faculty member
at Princeton's new Statistics Department. He held several visiting academic
appointments in the next decade at Harvard, UC Berkeley, University of
Texas at Austin and the University of Wisconsin at Madison. He was a full
professor at the University of Chicago in 1981--1983, and in 1984 moved back
to the Harvard Statistics Department, where he remains until now, and where
he served as chair from 1985 to 1994 and from 2000 to 2004.

Don has advised or coadvised over 50 Ph.D. students, written or edited 12
books, and published nearly 400 articles. According to Google Scholar, by
May 2014, Rubin's academic work has 150,000 citations, 16,000 in 2013
alone, placing him at the top with the most cited scholars in the world.

For his many contributions, Don has been honored by election to Membership
in the US National Academy of Sciences, the American Academy of Arts and
Sciences, the British Academy, and Fellowship in the American Statistical
Association, Institute of Mathematical Statistics, International
Statistical Institute, Guggenheim Foundation, Humboldt Foundation and
Woodrow Wilson Society. He has also received the Samuel S. Wilks Medal from
the American Statistical Association, the Parzen Prize for Statistical
Innovation, the Fisher Lectureship and the George W. Snedecor Award of the
Committee of Presidents of Statistical Societies. He was named Statistician
of the Year by the American Statistical Association's Boston and Chicago
Chapters. In addition, he has received honorary degrees from Bamberg
University, Germany and the University of Ljubljana, Slovenia.

Besides being a statistician, he is a music lover, audiophile and fan of
classic sports cars.

This interview was initiated on August 7, 2013, during the Joint
Statistical Meetings 2013 in Montreal, in anticipation of Rubin's 70th
birthday, and completed at various times over the following months.

\section*{Beginnings}

\textbf{Fan:} Let's begin with your childhood. I understand you grew up in
a family of lawyers, which must have heavily influenced you intellectually.
Can you talk a little about your family?

\textbf{Don:} Yes. My father was the youngest of four brothers, all of whom
were lawyers, and we used to have stimulating arguments about all sorts of
topics. Probably the most argumentative uncle was Sy (Seymour Rubin, senior
partner at Arnold, Fortas and Porter, diplomat, and professor of law at
American University), from D.C., who had framed personal letters of thanks
for service from all the presidents starting with Harry Truman and going
through Jerry Ford, as well as from some contenders, such as Adlai
Stevenson, and various Supreme Court Justices. I found this impressive but
daunting. The relevance of this is that it clearly created in me a deep
respect for the principles of our legal system, to which I find statistics
highly relevant---this has obviously influenced my own application of
statistics to law, for example, concerning issues as diverse as the death
penalty, affirmative action and the tobacco litigation.

\textbf{Fabri:} We will surely get back to these issues later, but was
there anyone else who influenced your interest in statistics?

\textbf{Don:} Probably the most influential was Mel, my mother's brother, a
dentist (then a bachelor). He loved to gamble small amounts, either in the
bleachers at Wrigley Field, betting on the outcome of the next pitch, while
watching the Cubs lose, or at Arlington Race track, where I was taught at a
young age how to read the Racing Form and estimate the ``true'' odds from
the various displayed betting pools, while losing two dollar bets.
Wednesday and Saturday afternoons, during the warm months when I was a
preteen, were times to learn statistics---even if at various bookie joints
that were sometimes raided. As I recall, I was a decent student of his, but
still lost small amounts.

There were two other important influences on my statistical interests from
the late 1950s and early 1960s. First, there was an old friend of my
father's from their government days together, a Professor Emeritus of
Economics at UC Berkeley, George Mehren, with whom I had many entertaining
and educational (to me) arguments, which generated a respect for economics
that continues to grow to this day. And second, my wonderful teacher of
physics at Evanston Township High School---Robert Anspaugh---who tried to
teach me to think like a real scientist, and how to use mathematics in the
pursuit of science.

By the time I left high school for college, I appreciated some statistical
thinking from gambling, some scientific thinking from physics, and I had
deep respect for disciplines other than formal mathematics, in particular,
physics and the law. These, in hindsight, are exposures that were crucial
to the kind of statistics to which I gravitated in my later years. More
details of the influence of my mentors can be found in Rubin (\citeyear{Rub14N2}).
%
\begin{figure}

\includegraphics{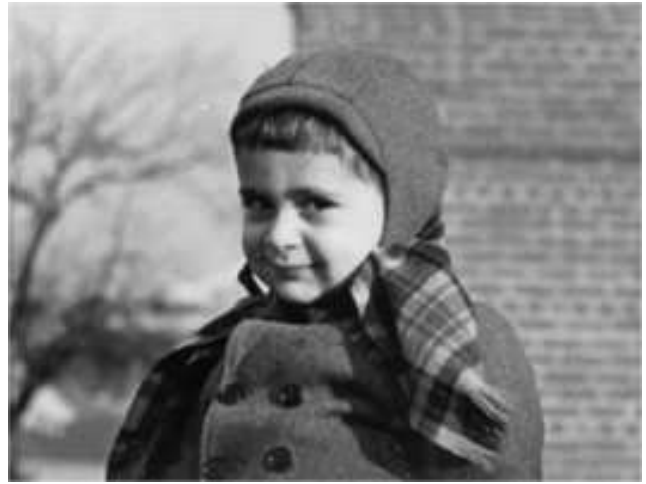}

\caption{Five-year old D. B. Rubin.}
\end{figure}

\section*{College Time at Princeton}

\textbf{Fan:} You entered Princeton in 1961, first as a physics major, but
later changed to psychology. Why the change and why psychology?

\textbf{Don:} That's a good question. Inspired by Anspaugh, I wanted to
become a physicist. I was lined up for a BA in three years when I entered
Princeton, and unknown to me before I entered, also lined up for a crazy
plan to get a Ph.D. in physics in five years, in a program being reconditely
planned by John Wheeler, a very well-known professor of physics there (and
Richard Feynman's Ph.D. advisor years earlier). In retrospect, this was a
wildly over-ambitious agenda, at least for me. For a combination of
complications, including the Vietnam War (and its associated drafts) and
Professor Wheeler's sabbatical at a critical time, I think no one succeeded
in completing a five-year Ph.D. from entry. In any case, there were many kids
like me at Princeton then, who, even though primarily interested in math
and physics, were encouraged to explore other subjects. I did that, and one
of the courses I took was on personality theory, taught by a wonderful
professor, Silvan Tomkins, who later became a good friend. At the end of my
second year, I switched from Physics\vadjust{\goodbreak} to Psychology, where my mathematical
and scientific background seemed both rare and appreciated---it was an
immature decision (not sure what a mature one would have been), but a fine
one for me because it introduced me to some new ways of thinking, as well
as to new fabulous academic mentors.

\textbf{Fabri:} You had some computing skills which were uncommon then,
right? So you started to use computers quite early.

\textbf{Don:} Yes. Sometime between my first and second year at Princeton,
I taught myself Fortran. As you mentioned, those skills were not common,
even at places like Princeton then.

\textbf{Fabri:} Was learning Fortran just a matter of having fun or did you
actually use these skills to solve problems?

\textbf{Don:} It was for solving problems. When I was in the Psychology
Department, I was helping to support myself by coding some of the early
batch computer packages for PSTAT, a Princeton statistical software
package, which competed with BMDP of UCLA at the time. I also wrote various
programs for simulating human behavior.

\textbf{Fan:} In your senior year at Princeton, you applied for Ph.D.
programs in psychology and were accepted by several very good places.

\textbf{Don:} Yes, I~was accepted by Stanford, Michigan and Harvard. I met
some extraordinary people during my visits to these programs. I went out to
Stanford first, and met William Estes, a quiet but wonderful professor with
strong mathematical skills and a wry wit, who later moved to Harvard.
Michigan had a very strong mathematical psychology program, and when I
visited in the spring of 1965, I~was hosted primarily by a very promising
graduating Ph.D. student, Amos Tversky, who was doing extremely interesting
work on human behavior and how people handled risks. In later years, he
connected with another psychologist, Daniel Kahneman, and they wrote a
series of extremely influential papers in psychology and economics, which
eventually led to Kahneman's winning the Nobel Prize in Economics in 2002;
Tversky passed away in 1996 and was thus not eligible for the Nobel Prize.
Kahneman (who recently was awarded a National Medal of Science by President
Obama) always acknowledges that the Nobel Prize was really a joint award
(to Tversky and him). I~was on a committee sometime last year with
Kahneman, and it was interesting to find out that I had known Tversky
longer than he had.

\textbf{Fan:} But ultimately you chose Harvard.

\textbf{Don:} Well, we all make strange decisions. The reason was that I
had an east-coast girlfriend who had another year in college.

\section*{Graduate Years at Harvard}

\textbf{Fabri:} You first arrived at Harvard in 1965 as a Ph.D. student in
psychology, which was in the Department of Social Relations then, but were
soon disappointed, and switched to computer science. What happened?

\textbf{Don:} When I visited Harvard in the summer of 1965, some senior
people in Social Relations appeared to find my background, in subjects like
math and physics, attractive, so they promised me that I could skip some of
the basic more ``mathy'' requirements. But when I arrived there, the chair
of the department, a sociologist, told me something like, ``No, no, I
looked over your transcript and found your undergraduate education
scientifically deficient because it lacked `methods and statistics'
courses. You will have to take them now or withdraw.'' Because of all the
math and physics that I'd had at Princeton, I felt insulted! I had to get
out of there. Because I had independent funding from an NSF graduate
fellowship, I looked around. At the time, the main applied math appeared
being done in the Division of Engineering and Applied Physics, which
recently became the Harvard's ``School of Engineering and Applied
Sciences.'' The division had several sections; one of them was computer
science (CS), which seemed happy to have me.

\textbf{Fan:} But you got bored again soon. Was this because you found the
problems in CS not interesting or challenging enough?

\textbf{Don:} No, not really that. There were several reasons. First, there
was a big emphasis on automatic language translation, because it was cold
war time, and it appeared that CS got a lot of money for computational linguistics from ARPA (Advanced
Research Projects Agency), now known as DARPA. The Soviet Union, from
behind the iron curtain, produced a huge number of documents in Russian,
but evidently there were not enough people in the US to translate them. A
complication is that there are sentences that you could not translate
without their context. I still remember one example: ``Time flies fast,''
a~three-word sentence that has three different meanings depending on which of
the three words is the verb. If this three-word sentence cannot be
automatically translated, how can one get an automatic (i.e., by computer)
translation of a complex paragraph? Related to this was Noam Chomsky's work
on transformational grammars, down the river at MIT.

Second, although I found some real math courses and the ones in CS on mathy
topics, such as computational complexity, which dealt with Turing machines,
Godel's theorem, etc., interesting, I found many of the courses dull. Much
of the time they were about programming. I remember one of my projects was
to write a program to project 4-dimensional figures into 2-dimensions, and
then rotate them using a DEC \mbox{PDP-1}. It took an enormous number of hours.
Even though my program worked perfectly, I felt it was a gigantic waste of
time. I also got a $\mathrm{C}+$ in that course because I never went to any
of the classes. Now, having dealt with many students, I would be more
sympathetic that I deserved a $\mathrm{C}+$, but not when I was a kid. At
that time, I~figured there must be something better to do than rotating 4D
objects and getting a $\mathrm{C}+$. But marching through rice paddies in
Vietnam or departing for somewhere in Canada didn't seem appealing. So
after picking up a MS degree in CS in 1966, although I stayed another year
in CS, I was ready to try something else.\looseness=-1

\textbf{Fabri:} How did statistics end up in your path?

\textbf{Don:} A summer job in Princeton in 1966 led to it. I did some
programming for John Tukey in Fortran, LISP and COBOL. I also did some
consulting for a Princeton sociology professor, Robert Althauser, basically
writing programs to do matched sampling, matching blacks and whites, to
study racial disparity in dropout rates at Temple University. I had a
conversation with Althauser about how psychology and then CS weren't
working out for me at Harvard. Because Bob was doing some semi-technical
things in sociology, he knew of Fred Mosteller, although not personally,
and also knew that Harvard had a decade-old Statistics Department that was
founded in 1957. He suggested that I contact Mosteller. After getting back
to Harvard, I~talked to Fred, and he suggested that I take some stat
courses. So in my third year in Harvard, I took mostly stat courses and did
OK in them. And the Stat department said ``Yes'' to me. It also helped to
have my own NSF funding, which I had from the start; they kept renewing for
some reason, showing their bad taste probably, but it worked out well for
me. Anyway, at the end of my third year at Harvard, I had switched to
statistics, my third department in four years.

\textbf{Fabri:} Besides Mosteller, who else was on the statistics faculty
then? It was a quite new department, as you said.

\textbf{Don:} The other senior people were Bill Cochran and Art Dempster,
who had recently been promoted to tenure. The junior ones were Paul
Holland; Jay Goldman, a probabilist; and Shulamith Gross from Berkeley, a
student of Erich Lehmann's.

\textbf{Fabri:} And you decided to work with Bill.

\textbf{Don:} Actually, I first talked to Fred. Fred always had a lot of
projects going; one was with John Tukey and he proposed that I work on it.
I told him that I had this matched sampling project of my own, and he
suggested that I talk to Cochran---Cochran a few years earlier was an
advisor for the Surgeon General's report on smoking and lung cancer. It was
obviously based on observational data, not on randomized experiments, and
Fred said that Cochran knew all about these issues in epidemiology and
biostatistics. So I went to knock on Bill's door. He answered with a grumpy\vadjust{\goodbreak}
sounding ``yes,'' I went in and he said, ``No, not now, later!'' So I
thought ``Hmmm, rough guy,'' but actually he was a sweetheart, with a great
Scottish dry sense of humor and a love of scotch and cigarettes (I
understand the former, although not the latter).

\textbf{Fabri:} Cochran did have a lasting influence on you, right?

\textbf{Don:} Yes, he had a tremendous influence on me. Once I was doing
some irrelevant math on matching, which I now see popping up again in the
literature. I~showed that to Bill, and he asked me, ``Do you think that's
important, Don?'' I said, ``Well, I don't know.'' Then he said, ``It is not
important to me. If you want to work on it, go find someone else to advise
you. I care about statistical problems that matter, not about making things
epsilon better.'' Another person who was very influential was Art Dempster.
Once I did some consulting for Data Text, a collection of batch computer
programs like PSTAT or BMDP. I was designing programs to calculate analyses
of variance, do regressions, ordinary least squares, matrix inversions, all
when you have, in hindsight, limited computing power. For advice on some of
those I talked to Dempster, who always has great multivariate insights
based on his deep understanding of geometry---very Fisherian.

\textbf{Fan:} Your Ph.D. thesis was on matching, which is the start of your
life-long pursuit of causal inference. How did your interest in causal
inference start?

\textbf{Don:} When I worked with Althauser on the racial disparity problem,
I~always\vadjust{\goodbreak} emphasized to him that it was inherently descriptive, not really
causal. I~remembered enough from my physics education in high school and
Princeton that association is not causation. So I was probably not
intrigued by causal inference per se, but rather by the confusion that the
social scientists had about it. You have to describe a real or hypothetical
experiment where you could intervene, and after you intervene, you see how
things change, not in time but between intervention (i.e., treatment)
groups. If you are not talking about intervention, you can't talk about
causality. For some reason, when I look at old philosophy, it seems to me
that they didn't get it right, whereas in previous centuries, some
experimenters got it. They bred cows, or mated hunting falcons. If you
mated excellent female and male falcons, the resulting next generation of
falcons would generally be better hunters than those resulting from random
mating. In the 20th century, many scientists and experimentalists got it.

\textbf{Fabri:} So you were only doing descriptive comparisons in your Ph.D.
thesis, and the notation of potential outcomes was not there.

\textbf{Don:} Partly correct. At that time, the notation of potential
outcomes was in my mind, because that is the way that Cochran initiated
discussions of randomized experiments in the class he taught in 1968.
Initially, it was all based on randomization, unbiasedness, Fisher's test,
etc. But the concepts had to be flipped into ordinary least squares (OLS)
regression and analysis of variance tables, because nobody could compute
anything difficult back then. One of the lessons in Bill's class in
regression and experimental design was to use the abbreviated Dolittle
method to invert matrices, by hand! So you really couldn't do randomization
tests in any generality. The other reason I was interested in experiments
and social science was my family history. There was always this legal
question lurking: ``But for this alleged misconduct, what would have
happened?''

\textbf{Fan:} What was your first job after getting your Ph.D. degree in
1970?

\textbf{Don:} I stayed at Harvard for one more year, as an instructor in
the Statistics Department, partly supported by teaching, partly supported
by the Cambridge Project, which was an ARPA funded Harvard--MIT joint
effort; the idea was to bring the computer science technologies of MIT and
the social sciences research of Harvard together to do wonderful things in
the social sciences. In the Statistics Department, I was coteaching with
Bob Rosenthal the ``Statistics for Psychologists''\vadjust{\goodbreak} course that, ironically,
the Social Relations Department wanted me to take five years earlier,
thereby driving me out of their department! Bob had, and has, tremendous
intuition for experimental design and other practical issues, and we have
written many things together.
%
\begin{figure}

\includegraphics{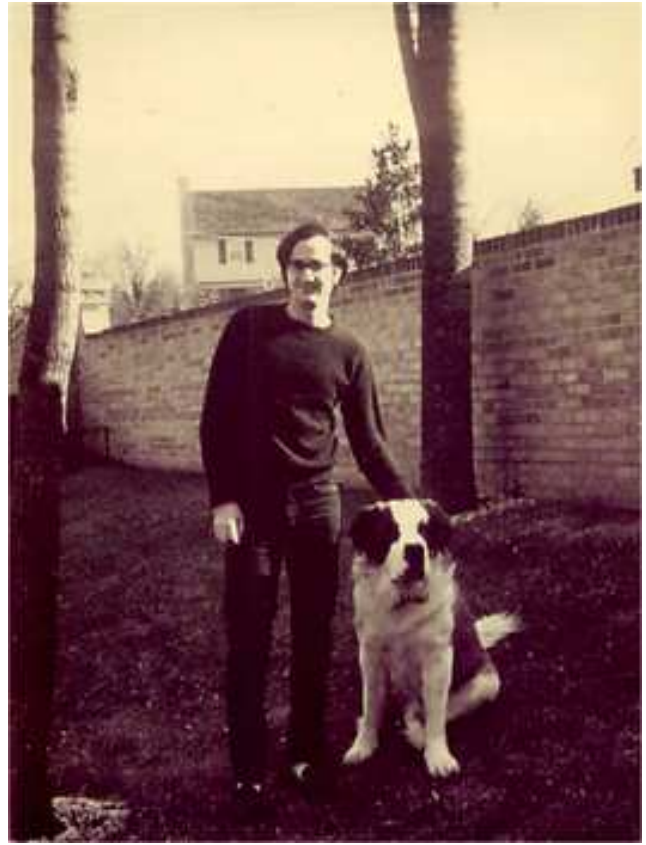}

\caption{D. B. Rubin (on left) with his puppy friend Thor (on right), about 1967.}
\end{figure}

\section*{The ETS Decade: Missing Data, EM and Causal Inference}

\textbf{Fan:} After that one year, you went for a position at ETS in
Princeton instead of a junior faculty position in a research university. It
was quite an unusual choice, given that you could probably have found a
position in a respected university statistics department easily.

\textbf{Don:} Right---many people thought I was goofy. I~did have several
good offers, one was to stay at Harvard, and another was to go to
Dartmouth. But I met Al Beaton, who was later my boss at ETS in Princeton,
at a conference in Madison, Wisconsin, and he offered me a job, which I
took. Al had a doctorate in education at Harvard, and had worked with
Dempster on computational issues, such as the ``sweep operator.'' He was a
great guy with a deep understanding of practical computing issues. Also, he
appreciated my research. Because I was an undergrad at Princeton, it was
almost like going home. For several years, I taught one course at
Princeton. Between the jobs at ETS and Princeton, I~was earning twice what
the Harvard salary would have been, which allowed me to buy a house on an
acre and a half, with a garage for rebuilding an older Mercedes roadster,
etc. A different style of life from that in Cambridge.

\textbf{Fan:} You seem to have had a lot of freedom to pursue research at
the ETS. What was your responsibility at ETS?

\textbf{Don:} The position at ETS was like an academic position with
teaching responsibilities replaced by consulting on ETS's social science
problems, including psychological and educational testing ones. I~found
consulting much easier for me than teaching, and ETS had interesting
problems. Also there were many very good people around, like Fred Lord, who
was highly respected in psychometrics. The Princeton faculty was great,
too: Geoffrey Watson (of the Durbin--Watson statistic) was the chair; Peter
Bloomfield was there as a junior faculty member before he moved to North
Carolina; and of course Tukey was still there, even though he spent a lot
of time at Bell Labs. John was John, having a spectacular but very unusual
way of thinking---obviously a genius. Stuart Hunter was in the Engineering
School then. These were fine times for me, with tremendous freedom to
pursue what I regarded as important work.

\textbf{Fabri:} By any measure, your accomplishments in the ETS years were
astounding. In 1976, you published the paper ``Inference and Missing Data''
in Biometrika (Rubin, \citeyear{Rub76}) that lays the foundation for modern analysis of
missing data; in 1977, with Arthur Dempster and Nan Laird, you published
the EM paper ``Maximum Likelihood from Incomplete Data via the EM
Algorithm'' in JRSS-B (Dempster, Laird and Rubin, \citeyear{DemLaiRub77}); in 1974, 1977,
1978, you published a series of papers that lay the foundation for the
Rubin Causal Model (Rubin, \citeyear{Rub74}, \citeyear{Rub77}, \citeyear{Rub78N1}). What was it like for you at
that time? How come so many groundbreaking ideas exploded in your mind at
the same time?

\textbf{Don:} Probably the most important reason is that I always worried
about solving real problems. I didn't read the literature to uncover a hot
topic to write about. I always liked math, but I never regarded much of
mathematical statistics as real math---much of it is just so tedious. Can
you keep track of these epsilons?

\textbf{Fabri:} There is no coincidence that all these papers share the
common theme of missing data.

\textbf{Don:} That's right. That theme arose when I was a graduate student.
The first paper I wrote on missing data, which is also my first
sole-authored paper, was on analysis of variance designs, a quite
algorithmic paper. It was always clear to me, from the experimental design
course from Cochran that you should set up experiments as missing data
problems, with all the potential outcomes under the not-taken treatments
missing. But nobody did observational studies that way, which seemed very
odd to me. Indeed, nobody was using potential outcomes outside the context
of randomized experiments, and even there, most writers dropped potential
outcomes in favor of least squares when actually doing things.

\textbf{Fan:} What was the state of research on missing data before you
came into the scene?

\textbf{Don:} It was extremely ad hoc. The standard approach to missing
data then was comparing the biases of filling in the means, or of
regression imputation under different situations, but almost always under
an implicit ``missing completely at random'' assumption. The purely
technical sides of these papers are solid. But I found there were always
counter examples to the propriety of the specific methods being considered,
and to explore them, one almost needed a master's thesis for each
situation. I would rather address the class of problems with some
generality. There is a mechanism that creates missing data, which is
critical for deciding how to deal with the missing data. That idea of
formal indicators for missing data goes way back in the contexts of
experimental design and survey design. I am consistently amazed how this
was not used in observational studies until I did so in the 1970s; maybe
someone did, but I've looked for years and haven't found anything. But
probably because the missing data paper was done in a relatively new way, I had great
difficulty in getting it published (more details in
Rubin, \citeyear{Rub14N1}).

\textbf{Fan:} The EM algorithm is another milestone in modern statistics;
it is also relevant in computer science and one of the most important
algorithm in data mining. Though similar ideas had been used in several
specific contexts before, nobody had realized the generality of EM. How did
Dempster, Laird and you discover the generality?

\textbf{Don:} In those early years at ETS, I had the freedom to remain in
close contact with the Harvard people, Cochran, Dempster, Holland and
Rosenthal, which was very important to me. I always enjoyed talking to
Dempster, who is a very principled and deep thinker. I was able to arrange
some consulting projects at ETS to bring him to Princeton. Once we were
talking about some missing data problem, and we started discussing filling
these values in, but I knew it wouldn't work in generality. I pointed to a
paper by Hartley and Hocking (\citeyear{HarHoc71}), where they deserted the approach of
iteratively filling in missing values, as in Hartley (\citeyear{Har56}) for the counted
data case, and went to Newton--Raphson, I~think, in the normal case. Even
though aspects of EM were known for years, and Hartley and others were sort
of nibbling around the edges of EM, apparently nobody put it all together
as a general algorithm. Art and I realized that you have to fill in
sufficient statistics. I had all these examples like t distributions,
factor analysis (the ETS guys loved that), latent class models. And Art had
a great graduate student, Nan Laird, available to work on parts of it, and
we started writing it up. The EM paper was accepted right away by JRSS-B,
even with invited discussions.

\textbf{Fan:} Now let's talk more about causal inference. You are known for
proposing the general potential outcome framework. It was Neyman who first
mentioned the notation of potential outcomes in his Ph.D. thesis (Neyman,
\citeyear{Spl90}), but the notation seemed to have long been neglected.

\textbf{Don:} Yes, it was ignored outside randomized experiments. Within
randomized studies, the notion became standard and used, for example, in
Kempthorne's work, but as I mentioned earlier, ignored otherwise.

\textbf{Fan:} Were you aware of Neyman's work before?

\textbf{Don:} No. I wasn't aware of his work defining potential outcomes
until 1990 when his Ph.D. thesis was translated into English, although I
attributed much of the perspective to him because of his work on surveys in
Neyman (\citeyear{Ney34}) and onward (see Rubin, \citeyear{Rub90N1}, followed by Rubin, \citeyear{Rub90N2}).

\textbf{Fabri:} You actually met Neyman when you visited Berkeley in the
mid-1970s. During all those lunches, had you ever discussed causal
inference and potential outcomes with him?

\textbf{Don:} I did. In fact, I had an office right next to his. Neyman
came to Berkeley in the late 30s. He was very impressive, not only as a
mathematical statistician, but also as an individual. There was a
tremendous aura about him. Shortly after arriving in Berkeley, I gave a
talk on missing data and causal inference. The next day, I went to lunch
with Neyman and I said something like, ``It seems to me that formulating
causal problems in terms of missing potential outcomes is an obvious thing
to do, not just in randomized experiments, but also in observational
studies.'' Neyman answered to the effect that (remarkable in hindsight
because he did so without acknowledging that he was the person who first
formulated potential outcomes), ``No, causality is far too speculative in
nonrandomized settings.'' He repeated something like this quote from his
biography, ``$\dots$Without randomization an experiment has little value
irrespective of the subsequent treatment.'' (Also see my comment on this
conversion in Rubin, \citeyear{Rub10}.) Then he went to say politely but firmly,
``Let's not talk about that, let's instead talk about astronomy.'' He was
very into astronomy at the time.

\textbf{Fabri:} You probably learned the reasons why he was so involved in
the frequentist approach.

\textbf{Don:} Yes. I remember we once had a conversation about what
confidence intervals really meant and why the formal Neyman--Pearson
approach seemed irrelevant to me. He said something like, ``You
misinterpret what we have done. We were doing the mathematics; go back and
read my 1934 paper where I first defined a confidence interval.'' He
defined it as a procedure that has the correct coverage for all prior
distributions (see page 589, Neyman, \citeyear{Ney34}). If you think of that, you are
forced to include all point mass priors and, therefore, you are forced to
do Neyman--Pearson. He went on to say (approximately), ``If you are a real
scientist with a class of problems to work on, you don't care about all
point-mass priors, you only care about the priors for the class of problems
you will be working on. But if you are doing the mathematics, you can't
talk about the problems you or anyone is working on.'' I tried to make this
point in a comment (Rubin, \citeyear{Rub95}), but it didn't seem to resonate to
others.

\textbf{Fabri:} In his famous 1986 JASA paper, Paul Holland coined the term
``Rubin Causal Model (RCM),'' referring to the potential outcome framework
to causal inference (Holland, \citeyear{Hol86}). Can you explain why, if you think so,
the term ``Rubin Causal Model'' is a fair description of your contribution
to this topic?

\textbf{Don:} Actually Angrist, Imbens and I had a rejoinder in our 1996
JASA paper (Angrist, Imbens and Rubin, \citeyear{AngImbRub96}), where we explain why we think
it is fair. Neyman is pristinely associated with the development of
potential outcomes in randomized experiments, no doubt about that. But in
the 1974 paper (Rubin, \citeyear{Rub74}), I made the potential outcomes approach for
defining causal effects front and center, not only in randomized
experiments, but also in observational studies, which apparently had never
been done before. As Neyman told me back in Berkeley, in some sense, he
didn't believe in doing statistical inference for causal effects outside of
randomized experiments.

\textbf{Fan:} Also there are features in the RCM, such as the definition of
the assignment mechanism, that belong to you.

\textbf{Don:} Yes, it was crucial to realize that randomized experiments
are embedded in a larger class of assignment mechanisms, which was not in
the literature. Also, in the 1978 paper (Rubin, \citeyear{Rub78N1}), I proposed three
integral parts to this RCM framework: potential outcomes, assignment
mechanisms, and a (Bayesian) model for the science (the potential outcomes
and covariates). The last two parts were not only something that Neyman
never did, he possibly wouldn't even like the third part. In fact, I think
it is unfair to attribute something to someone who is dead, who may not
approve of the content being attributed. If the fundamental idea is clear,
such as with Fisher's randomization test of a sharp null hypothesis, sure,
attribute that idea to Fisher no matter what the test statistic, as in
Brillinger, Jones and Tukey (\citeyear{Bri}). Panos Toulis (a~fine Harvard Ph.D.
student) helped me track down this statement that I remembered reading in
my ETS days from a manuscript John gave to me:

``\textit{In the precomputer era, the fact that almost all work could be
done once and for all was of great importance. As a consequence, the
advantages of randomization approaches---except for those few cases where
the randomization distributions could be dealt with once and for all---were not adequately valued}.

\textit{One reason for this undervaluation lay in the fact that, so long as
randomization was confined to specially manageable key statistics, there
seemed no way to introduce into the randomization approach the
insights---some misleading and some important and valuable---into what test
statistics would be highly sensitive to the changes that it was most
desired to detect. The disappearance of this situation with the rise of the
computer seems not to have received the attention that it deserves}.''
(Brillinger, Jones and Tukey, \citeyear{Bri}, Chapter~25, page F-5.)

\textbf{Fabri:} Here I am quoting an interesting question by Tom Belin
regarding potential outcomes: ``Do you believe potential outcomes exist in
people as fixed quantities, or is the notion that potential outcomes are a
device to facilitate causal inference?''

\textbf{Don:} Definitely the latter. Among other things, a~person's
potential outcomes could change over time, and how do we know the people
who were studied in the past are still exchangeable with people today? But
there are lots of devices like that in science.

\textbf{Fan:} In the RCM, cause/intervention should always be defined
before you start the analysis. In other words, the RCM is a framework to
investigate the ``effects of a cause,'' but not the ``causes of an
effect.'' Some criticize this as a major limitation. Do you regard this as
a limitation? Do you think it is ever possible to draw inference on the
causes of effects from data, or is it, per se, an interesting
question worth further investigation?

\textbf{Don:} I regard ``the cause'' of an event topic as more of a
cocktail conversation topic than a scientific inquiry, because it leads to
an essentially infinite regress. Someone says, ``He died of lung cancer
because he smoked three packs a day''; then someone else counters, ``Oh no,
he died of lung cancer because both of his parents smoked three packs a day
and, therefore, there was no hope of his doing anything other than smoking
three packs a day''; then another one says, ``No, no, his parents smoked
because his grandparents smoked---they lived in North Carolina where, back
then, everyone smoked three packs a day, so the cause is where the
grandparents lived,'' and so on. How far back should you go? You can't talk
sensibly about \textit{the cause} of an event; you can talk about ``but for
that cause (and there can be many `but for's), what would have happened?''
All these questions can be addressed hypothetically. But \textit{the
cause}? The notion is meaningless to me.

\textbf{Fabri:} Do you feel that you benefit from knowing about history of
statistics when you are thinking about fundamentals of statistics?

\textbf{Don:} I know some history, but not a large amount. The most
important lessons occur when I feel that I understand why one of the
giants, like Fisher or Neyman, got something wrong. When you understand why
a mediocre thinker got something wrong, you learn little, but when you
understand why a genius got something wrong, you learn a tremendous
amount.
%
\begin{figure}

\includegraphics{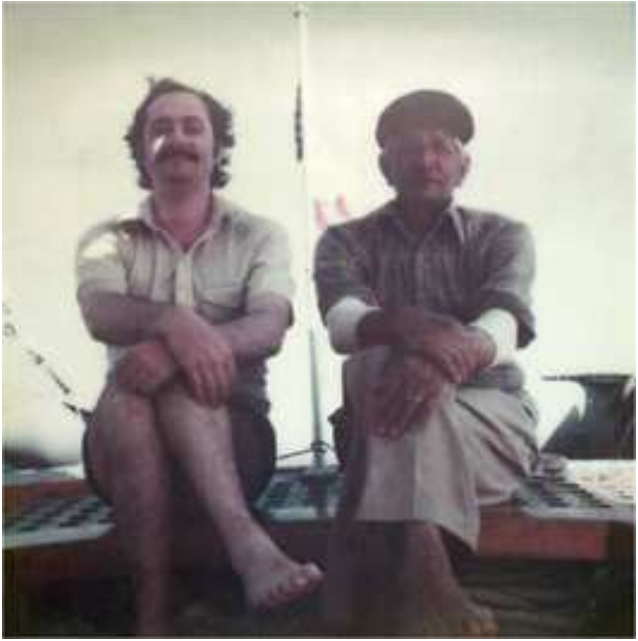}

\caption{D. B. Rubin (on left) poses with the captain (on right) of Sy's boat
harbored in Bodrum, Turkey, mid-1970s.}
\end{figure}

\section*{Back to Harvard: Propensity Score, Multiple Imputation and More}

\textbf{Fabri:} After those productive years at ETS, you spent some time at
the EPA (US Environmental Protection Agency). Why did you decide to move,
given that you were apparently doing very well at the ETS?

\textbf{Don:} It started partly from my joking answer to the question,
``How long have you been at ETS?'' I answered, ``Too long.'' The problems
that I had dealt with at ETS started to appear repetitive, and I felt that
I had made important contributions to them including EM and multiple
imputation ideas, which were being used to address some serious issues,
like test equating, and formulating the right ways to collect data. So I
wanted to try something else. At the time, David Rosenbaum was the head of
the Office of Radiation Programs at the EPA. He had the grand idea of
putting together a team of applied mathematicians and statisticians.
Somehow he found my name and invited me to D.C. to find out whether I
wanted to lead such a group. Basically, I had the freedom to hire several
people of my choice, and I had a good government salary (at the level of
``Senior Executive Service''). So I said, ``Let's see whom I can get.'' I
was able to convince both Rod Little (who was in England at that time) and
Paul Rosenbaum (whom I advised while I was still at ETS), as well as Susan
Hinkins, who wrote a thesis on missing data at Montana State University,
and two others. That was shortly before the presidential election. Then the
Democrats lost and Reagan was to come in, and everything seemed to be
falling apart. All of a sudden, many of the people above my level at the
EPA (most of whom were presidential appointments), had to prepare to turn
in their resignations, and had to be concerned about their next
positions.

\textbf{Fabri:} So the EPA project ended before it even got started.

\textbf{Don:} It didn't start at all in some sense. I formally signed on at
the beginning of December, and after one pay period, I turned in my
resignation. But I felt responsible to find jobs for all these people I
brought there. Eventually, Susan Hinkins got connected with Fritz Scheuren
at the IRS; Paul Rosenbaum got a position at the University of Wisconsin at
Madison; Rod got a job related to the Census. One nice thing about that
short period of time is that, through the projects I was in charge of, I
made several good connections, such as to Herman Chernoff and George Box.
George and I really hit it off, primarily because of his insistence on
statistics having connections to real problems, but also because of his
wonderful sense of humor, which was witty and ribald, and his love of good
spirits. In any case, the EPA position led to an invitation to visit Box at
the Math Research Center at the University of Wisconsin, which I gladly
accepted. That gave me the chance to finish writing the propensity score
papers with Paul (Rosenbaum and Rubin, \citeyear{RosRub83}, \citeyear{autokey27}, \citeyear{autokey28}).

\textbf{Fan:} Since you mentioned propensity score, arguably the most
popular causal inference technique in a wide range of applied disciplines,
can you give some insights on the ``natural history'' of propensity
score?

\textbf{Don:} I first met Paul in 1978, when I came to Harvard on a
Guggenheim fellowship; he was a first-year Ph.D. student, extremely bright
and devoted. Back in my Princeton days I did some consulting for a
psychologist at Rutgers, June Reinisch, who later became the first director
of the Kinsey Institute after Kinsey. She was very interested in studying
the nature-nurture controversy---what makes men and women so different?
She and her husband, who was also a psychologist, were doing experiments on
rats and pigs. They injected hormones into the uteri of pregnant animals,
and thereby exposed the fetuses to different prebirth environments; this
kind of randomized experiment is obviously unethical to do with humans. One
of the problems Paul and I were working on for this project, also as part
of Paul's thesis, was matching---matching background characteristics of
exposed and unexposed. The covariates included a lot of continuous and
discrete variables, some of which were rare events like certain serious
diseases prior to, or during, early pregnancy. Soon it became clear that
standard matching approaches, like Mahalanobis matching, do not work well
in such high dimensional settings. You have to find some type of summaries
of these variables and balance the summaries in the treatment and control
groups, not individual to individual. Then we realized if you have an
assignment mechanism, you can match on the individual assignment
probabilities, which is essentially the Horvitz--Thompson idea, to
eliminate all systematic bias. I don't remember the exact details, but I
think we first got the propensity score idea when working on a Duke data
bank on coronary artery bypass surgery, but refined it for the Reinisch
data, which is very similar in principle. Again, the idea of the propensity
score is motivated by addressing real problems, but with generality.

\textbf{Fan:} Multiple Imputation (MI) is another very influential
contribution of yours. Your book ``Multiple Imputation for Nonresponse in
Sample Surveys'' (Rubin, \citeyear{Rub87N1}) has commonly been cited as the origin of
MI. But my understanding is that you first developed the idea and coined
the term much earlier.

\textbf{Don:} Correct, I first wrote about MI in an ASA proceedings paper
in 1978 (Rubin, \citeyear{Rub72}, \citeyear{Rub78N2}). That's where ``the $18+$ years'' comes from when I
wrote ``Multiple Imputation After $18+$ Years'' (Rubin, \citeyear{Rub96}).

\textbf{Fabri:} MI has been developed in the context of missing data, but
it applicability seems to be far beyond missing data.

\textbf{Don:} Yes, MI has been applied and will be, I~think, all over
the place. The reason I titled the book that way, ``Multiple Imputation for
Nonresponse in Sample Surveys,'' is that it was obvious to me that in the
settings where you need to create public-use data sets, you had to have a
separation between the person who fixed up the missing data problem and the
many people who might do analyses of the data. So there was an obvious need
to do something like this, because users could not possibly have the
collection of tools and resources to do the imputation, for example, using
confidential information. My Ph.D. students, Trivellore Raghunathan (Raghu)
and Jerry Reiter, have made wonderful contributions to confidentiality
using MI. Of course, other great Ph.D. students of mine Nat Schenker, Kim
Hung Lee, Xiao-Li Meng, Joe Schafer, as well as many others, have also made
major contributions to MI. The development of MI really reflects the
collective efforts from these people and others like Rod Little and his
colleagues and students.

\textbf{Fabri:} Rod Little once half-jokingly said, ``Want to be highly
cited? Coauthor a book with Rubin!'' And indeed he wrote the book
``Statistical Analysis with Missing Data'' with you (Little and Rubin,
\citeyear{LitRub87}, \citeyear{LitRub02}), which is now regarded as the classic textbook on missing data.
There have been a lot of new advances and changes in missing data since
then. Will we see a new edition of the book that incorporates these
developments sometime soon?

\textbf{Don:} Oh yes, we are working on that now. The main changes from
1987 to 2002 reflect the greater acceptability of Bayesian methods and MCMC
type computations. Rod is a fabulous coauthor, a much more fluid writer
than I am. I believe this third edition will have even more major changes
than the 2002 one did from the 1987 one, but again many driven by
computational advances.
%
\begin{figure}

\includegraphics{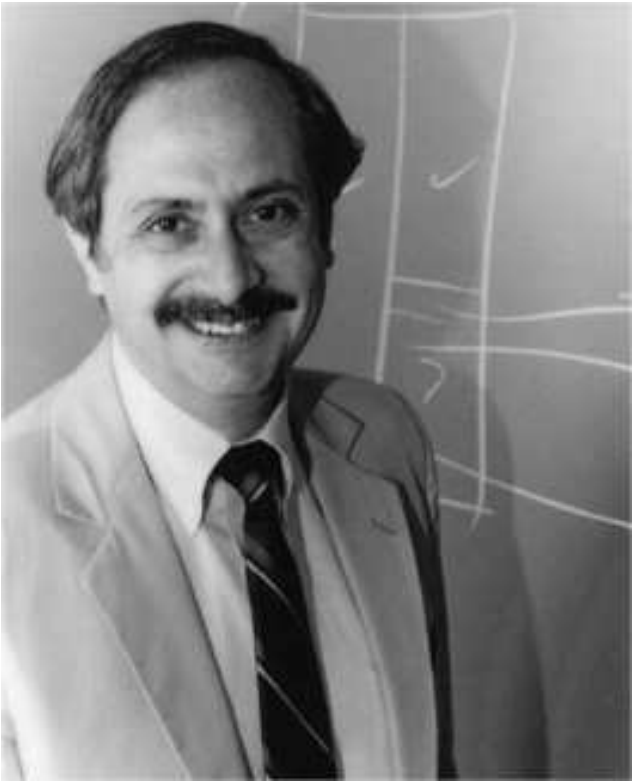}

\caption{D. B. Rubin at Harvard, early 1980s.}
\end{figure}

\section*{On Bayesian}

\textbf{Fan:} In the 1978 Annals paper (Rubin, \citeyear{Rub78N1}), you gave, for the
first time, a rigorous formulation of Bayesian inference for causal
effects. But the Bayesian approach to causal inference did not have much
following until very recently, and the field of causal inference is still
largely frequentist. How do you view the role of Bayesian approach in
causal inference?

\textbf{Don:} I believe being Bayesian is the right way to approach things,
because the basic frequentist approach, such as the Fisherian tests and
Neyman's unbiased estimates and confidence intervals, usually does not work
in complicated problems with many nuisance unknowns. So you have to go
Bayesian to create procedures. You can go partially Bayesian using things
like posterior predictive checks, where you put down a null that you may
discover evidence against, or direct likelihood approaches as in Frumento
et al. (\citeyear{Fruetal12}); if the data are consistent with a null that is interesting,
you live with it. But Neyman-style frequentist evaluations of Bayesian
procedures are still relevant.

\textbf{Fan:} But why is the field of causal inference still predominantly
frequentist?

\textbf{Don:} I think there are several reasons. First, there are many
Bayesian statisticians who are far more interested in MCMC algebra and
algorithms, and do not get into the science. Second, I regard the method of
moments (MOM) frequentist approach as pedagogically easier for motivating
and revealing sources of information. Take the simple instrumental variable
setting with one-sided noncompliance. Here, it is very easy to look at the
simple MOM estimate to see where information comes from. With Bayesian
methods, the answer is, in some sense, just there in front of you. But when
you ask where the information comes from, you have to start with any value,
and iterate using conditional expectations, or draws from the current joint
distributions. You have to have far more sophisticated mathematical
thinking to understand fully Bayesian ideas. There are these problems with
missing data (as in my discussion of Efron, \citeyear{Efr94}) where there are unique,
consistent estimates of some parameters using MOM, but for which the joint
MLE is on the boundary. So I think it is often easier, pedagogically, to
motivate simple estimators and simple procedures, and not try to be
efficient when you convey ideas. In causal inference, that corresponds to
talking about unbiased or nearly unbiased estimates of causal estimands, as
in Rubin (\citeyear{Rub77}). There are other reasons having to do with the current
education of most statisticians.

\textbf{Fan:} After EM, starting from the early 1980s, you were heavily
involved in developing methods for Bayesian computing, including the
Bayesian bootstrap (Rubin, \citeyear{Rub81}), the sampling importance-resampling (SIR)
algorithm (Rubin, \citeyear{Rub87N2}), and (lesser-acknowledged) ``approximate Bayesian
computation (ABC)'' (Rubin, \citeyear{Rub84}, Section~3.1).

\textbf{Don:} It was clear then that computers were going to allow Bayes to
work far more broadly than earlier. You, as well as others such as Simon
Tavare, Christian Robert and Jean-Michel Marin, are giving me credit for
first proposing ABC. Thanks! Although, frankly, I~never thought that would
be a useful algorithm except in problems with simple sufficient
statistics.

\textbf{Fabri:} But you do not seem to have followed up much on these ideas
later, even if you have used them. Also you do not label yourself as a
Bayesian or a frequentist, even if all these papers made extraordinary
contributions to Bayesian inference with fundamental and big ideas.

\textbf{Don:} First of all, fundamentally I am hostile to all
``religions.'' I recently heard a talk by Raghu in Bamberg, Germany, where
he said that in his world they have zillions of gods, and I think that is
right; you should have zillions of gods, one for this good idea, one for
that good idea. And different people can create different gods to whatever
extent they want to. I am not a fully-pledged member of the Bayesian
camp---I like being friends with them, but I never want to be religiously
Bayesian. My attitude is that any complication that creates problems for
one form of inference creates problems for all forms of inference, just in
different ways. For example, the fact that confounded treatment assignments
cause problems for frequentist inference is obvious. Does it generate
problems for the Bayesian? Yeah, that point was made in the 1978 Annals
paper: Randomization matters to a Bayesian, although not in the same way as
to a frequentist, that is, not as the basis for inference, but it affects
the likelihood function.

There is something I am currently working on with a Ph.D. student, Viviana
Garcia, that builds on a paper I wrote with Paul Rosenbaum in 1984
(Rosenbaum and Rubin, \citeyear{autokey29}), which is the only Bayesian paper that Paul
has ever written, at least with me. In that paper, we did some simulations
to show there is an effect on Bayesian inference of the stopping rule. We
show that if you have a stopping rule and use the ``wrong'' prior to do the
analysis, like a uniform improper prior, but the data are coming from a
``correct'' prior, and you look at the answer you get from the right prior
and from the ``wrong'' prior, they are different. The portion of the right
posterior that you cover using the ``wrong'' posterior is incorrect. This
extends to all situations and it is related to all of these ignorability
theorems, and it means that you need to have the right model with respect
to the right measure. Of course achieving this is impossible in practice
and, therefore, leads to the need for frequentist (Neymanian) evaluations
of the operating characteristics of Bayesian procedures when using
incorrect models (Rubin, \citeyear{Rub84}). Bayes works, in principle, there is no
doubt, but it can be so hard! It can work, in practice, but you must have
some other principles floating around somewhere to evaluate the
consequences---how wrong your conclusions can be. So you must have
something to fall back on, and I think that is where these frequentist
evaluations are extremely useful, not the unconditional Neyman--Pearson
frequentist evaluations for all point mass priors (which were critical as
mathematical demonstrations that we cannot achieve the ideal goal in any
generality), but evaluations for the class of problems that you are dealing
with in your situation.

\textbf{Fan:} The 1984 Annals paper ``Bayesianly Justifiable and Relevant
Frequency Calculations for the Applied Statistician'' (Rubin, \citeyear{Rub84}) is one
of my all-time favorite papers. This paper, as the earlier paper by George
Box (Box, \citeyear{Box80}), deals with the ``calibrated Bayes'' paradigm with
generality, which can be viewed as a compromising or mid-ground between the
Bayesian and frequentist paradigms. It has a profound influence on many of
us. In particular, Rod Little has strongly advocated ``calibrated Bayes''
as the 21st century roadmap of statistics in several of his
prominent talks, including the 2005 ASA President's Invited Address and the
2012 Fisher Lecture. What was the background and reasons for you to write
that paper?

\textbf{Don:} Interesting question. I was visiting Box at the Mathematics
Research Center in 1981--1982 and wrote Rubin (\citeyear{Rub83}) partly during that
period---I~think it's a good paper with some good ideas, but without a
satisfying big picture. That dissatisfaction led to that 1984 paper---what
is the big picture? It took me a very long time to ``get it right,'' but it
all seems very obvious to me now. The idea of posterior predictive checks
has been further articulated and advanced in Meng (\citeyear{Men94}), Gelman, Meng and
Stern (\citeyear{G96}), and the multiauthored book ``Bayesian Data Analysis'' (Gelman
et al., \citeyear{Geletal95}, \citeyear{Geletal03}, \citeyear{Geletal14}).

\textbf{Fabri:} Can you talk a little more about the ``Bayesian Data
Analysis'' book, probably one of the most popular Bayesian textbooks?

\textbf{Don:} Yup, I think that the Gelman et al. book might be THE most
popular Bayesian text. It started out as notes by John Carlin for a
Bayesian course that he taught when I was Chair sometime in the mid or late
1980s. Andy must have been a Ph.D. student at that time, with tremendous
energy for scholarship. John was heading back to Australia, which is his
homeland, and somehow the department had some extra teaching money, and we
wanted to keep John around for a year---I do not remember the details. But
I do remember that the idea of turning the notes for the course into a full
text was percolating. Also Hal Stern was an Associate Professor with us at
that time, and so the four of us decided to make it happen. We basically
divided up chapters and started writing. Even though John's initial notes
were the starting basis, things changed as soon as Andy ``took charge.''
Quickly, Andy and Hal were the most active. Andy, with Hal, were even more
dominant in the second edition, where I added some parts, edited others,
but clearly this was Andy's show. The third edition, which just came out in
early 2014, was even more extreme, with Andy adding two coauthors (David
Dunson and Aki Vehtari) because he liked their work, and they had been
responsive to Andy's requests. As the old man of the group, I just
requested that I be the last author; Andy obviously was the first author,
and the second and third were as in the first edition. In some ways, I feel
like I'm an associate editor of a journal that has Andy as the editor! We
get along fine, and clearly it's a successful book.

\textbf{Fan:} A revolutionary development in statistics since the early 90s
was the MCMC methodology. You left your mark in this with Gelman, proposing
the Gelman--Rubin statistic for convergence check (Gelman and Rubin, \citeyear{GelRub92}),
which seems to be very much connected to some of your previous work.

\textbf{Don:} Correct. We embedded the convergence check problem into the
combination of the multiple imputation and multiple chains frameworks,
using the idea of the combining rules for MI. The idea of using multiple
chains---that comes from physics---and was Andy's knowledge, not mine. My
contribution was to suggest using modified MI combining rules to help do
the assessment of convergence. The idea is powerful because it is so
simple. If the starting value does not matter, which is the whole point,
then it doesn't matter, period. The real issue should be how you choose the
functions of the estimands that you are assessing, and as always, you want
convergence to asymptotic normality to be good for these functions, so that
the simple justification for the Gelman--Rubin statistic is roughly
accurate.

\section*{The 1990s: Collaborating with Economists}

\textbf{Fabri:} In the 1990s, you started to work with economists. With
Joshua Angrist, and particularly with Guido Imbens, you wrote a series of
very influential papers, connecting the potential outcomes framework to
causal inference with instrumental variables. Can you tell us how this
collaboration started?

\textbf{Don:} Absolutely. I always liked economics; many economists are
great characters! It was in the early 90s when Guido came to my office as a
junior faculty member in the Harvard Economics Department and basically
said, ``I think I have something that may interest you.'' I had never met
him before, and he was asking if the concept of instrumental variables
already had a history in
%
\begin{figure*}

\includegraphics{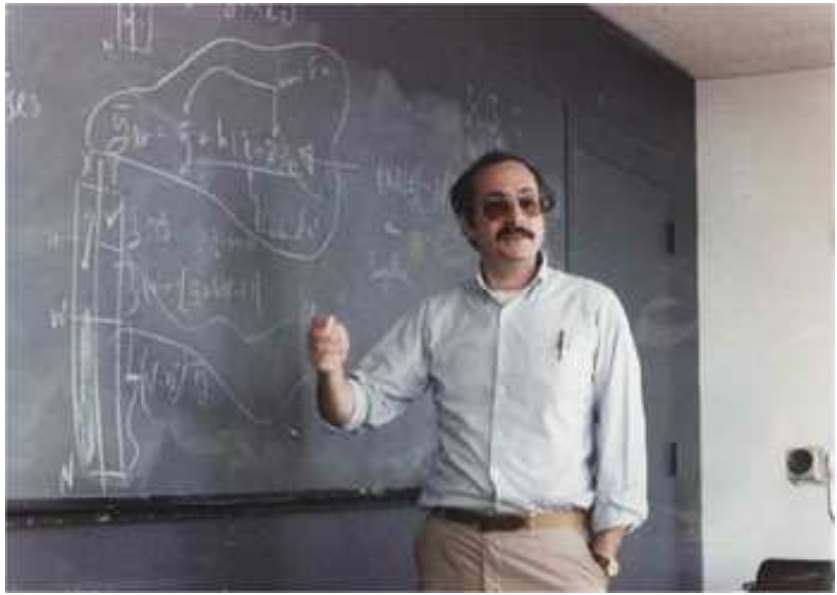}

\caption{In classroom at Harvard, late 1980s.}
\end{figure*}
statistics. Guido and Josh Angrist had already defined the LATE (local
average treatment effect) in an Econometrica paper (Imbens and Angrist,
\citeyear{ImbAng94})---although I think CACE (Complier Average Causal Effect) is a much
better name because it is more descriptive and more precise---local can be
local for anything, local for Boston, local for females, etc. Then I asked
in return, ``Well tell me the setup, I have never heard of it in statistics
before'' and while he was explaining I started thinking, ``Gosh, there is
something important here! I have never seen it before,'' and then I said,
``Let's meet tomorrow and talk about it more,'' because these kinds of
assumptions (monotonicity and the ``exclusion restriction'') were
fascinating to me, and it was clear that there was something there that I
had never really thought hard about; it was great. That eventually led to
the instrument variables paper (Angrist, Imbens and Rubin, \citeyear{AngImbRub96})
and the later Bayesian paper (Imbens and Rubin,
\citeyear{ImbRub97}).

A closely related development was a project I was consulting on for AMGEN
at about the same time, for a product for the treatment of ALS (amyotrophic
lateral sclerosis), or Lou Gehrig's disease, which is a progressive
neuromuscular disease that eventually destroys motor neurons, and death
follows. The new product was to be compared to the control treatment where
the primary outcome was quality of life (QOL) two years post-randomization,
as measured by ``forced vital capacity'' (FVC), essentially, how big a
balloon you can blow up. In fact, many people do not reach the end-point of
two-year post-randomization survival, and so two-year QOL is ``truncated''
or ``censored'' by death. People were trying to fit this problem into a
``missing data'' framework, but I realized right away that it was something
different.

\textbf{Fan:} Essentially both ideas are special cases of the general idea
of Principal Stratification, which we can discuss in a moment.

\textbf{Don:} Yes, indeed. These meetings with Guido and this way of
thinking were so much more articulated and close to the thinking of European
economists in the 30s and 40s, like Tinbergen and Haavelmo, than many
subsequent economists who seemed sometimes to be too into their OLS algebra
in some sense. There was some correspondence between one of the
two---Haavelmo, I think---and Neyman on these hypothetical experiments on
supply and demand. European brains were talking to each other, and not
simply exchanging technical mathematics!

\textbf{Fabri:} I know that many years before you met Guido, with other
statisticians, like Tukey, you had discussions about the way economists
were treating selection problems, or missing data problems. But you had
some adventurous, to say the least, previous experiences with economists
dealing with problems that you had worked on, which they had almost
neglected completely.

\textbf{Don:} Yes, James Heckman was tracking my work in the early 1980s
when I came to Chicago after ETS. The public exchange came out in the ETS
volume edited by Howard Wainer (which is where Glynn, Laird and Rubin,
\citeyear{GlyLaiRub86}, appears), with comments from Heckman, Tukey, Hartigan and others.

\textbf{Fabri:} Economics is a field where the idea of causality is
crucial; did you find interest in economics also for this very reason? The
problems they have are usually very interesting.
%
\begin{figure*}[b]

\includegraphics{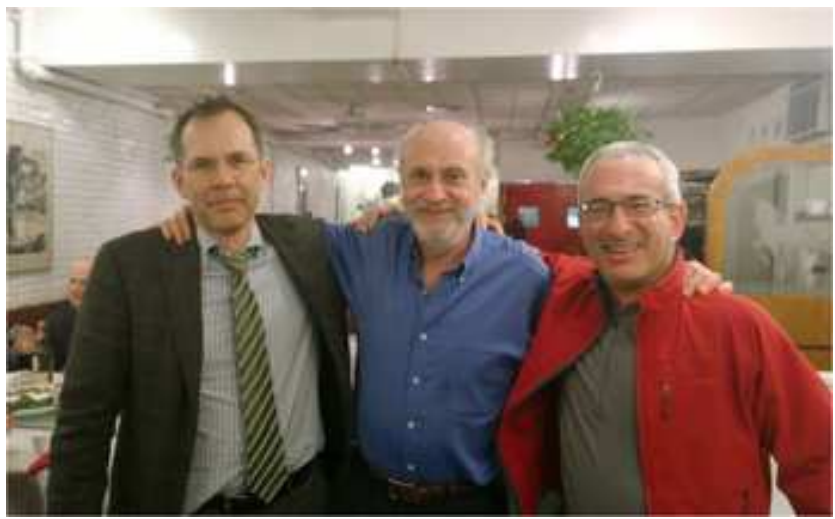}

\caption{(Left to right) Guido Imbens, Don Rubin, Josh Angrist. March,
2014.}
\end{figure*}

\textbf{Don:} There are often interesting questions from social science
students that come up in class. One recent example is how do we answer
questions like ``What would the Americas be like if they were not settled
by Europeans?'' I asked the questioner, ``Who would they be settled by
instead? By the Chinese? By the Africans? What are you talking about? What
are we comparing the current American world to?'' Another example comes
from an undergraduate thesis that I directed, by Alice Xiang, which won
both the Hoopes Prize and the economics' Harris Prize for an outstanding
honors thesis. The thesis is on the causal effect of racial affirmative
action in law school admissions on some outcomes versus the same proportion
of affirmative action admissions but counter-factually based on
socioeconomic status. This is not just for cocktail conversation---it was
a case recently before the US Supreme Court, Fisher v. University of Texas,
which was kicked back to the lower court to reconsider, and additionally
the issue was recently affected by a state law in Michigan. There is an
amicus brief sent to the US Supreme Court to which Guido (Imbens), former
Ph.D. students, Dan Ho, Jim Greiner and I (with others) contributed.

Such careful formulation of questions is something critical, and to me is
central to the field of statistics. It is crucial to formulate clearly your
causal question. What is the alternative intervention you are considering,
when you talk about the causal effect of affirmative action on graduation
rates or bar-passage rates? Immediately formulating the problem as an OLS
regression is the wrong way to do this, at least to me.

\textbf{Fan:} You apparently have a long interest in law; besides the
aforementioned ``affirmative action'' thesis, you have done some
interesting work in applied statistics in law.

\textbf{Don:} Yes. Paul Rosenbaum was, I think, the first of my Harvard
students who did something about statistics in law. Either his qualifying
paper or a class paper in 1978 was on the effect of the death penalty. Jim
Greiner, another great Ph.D. student of mine, who had a law degree before
entering Harvard Statistics, wrote his Ph.D. thesis (and subsequently several
important papers) on potential outcomes and causal effects of immutable
characteristics. He is now a full professor at the Harvard Law School.
There were also several previous undergraduate students of mine who were
interested in statistics and law, but (sadly) most went to law school.
Since 1980, I have been involved in many legal topics.

\section*{The New Millennium: Principal Stratification}

\textbf{Fabri:} The work you did with Guido, as well as the work on
censoring due to death, led to your paper on Principal Stratification
(Frangakis and Rubin, \citeyear{FraRub02}), coauthored with this brilliant student of
yours, Constantine Frangakis, who happens to be Fan's advisor.

\textbf{Don:} Yes, Constantine is fabulous, but the original title of that
paper was very long, same with the title of his thesis. It went on and on,
with probably a few Latin, a few Italian, a few French and a few Greek
words! Of course I was exasperated, so I convinced him to simplify the
paper's title to ``Principal Stratification in Causal Inference.'' He is
brilliant, so good that he has no trouble dealing with all the complexity
in his own mind, but therefore he struggles at times pulling out the
kernels of all these ideas, making them simple.

\textbf{Fan:} What do you think is the most remarkable thing about the
development of Principal Stratification?

\textbf{Don:} It is a whole new collection of ways of thinking about what
the real information is in causal problems. Once you understand what the
real information is, you can start thinking about how you can get the
answers to questions that you want to extract from that information; you
always have to make assumptions, and it forces you to explicate what these
assumptions are, not in terms of OLS, which no social scientist or doctor
would really understand---but in terms of scientific or medical entities.
And because you have to make assumptions, be honest and state them clearly.
For example, I like your papers (Mealli and Pacini, \citeyear{MeaPac13}; Mattei, Li and
Mealli, \citeyear{MatLiMea13}) about multiple post-randomization outcomes, where you discuss
that for some outcomes, exclusion restriction or other structural
assumptions may be more plausible.

\textbf{Fabri:} Principal Stratification is sometimes compared to other
tools for doing so-called mediation analysis---what is your view about
inferring on mediation effects?

\textbf{Don:} I think we (Don and Fabri) discussed a paper recently in
JRSS-A, and those discussions summarize my--our view on that. Essentially,
some of the people writing about mediation seem to misunderstand what a
function is. They write down something that has two arguments inside
parenthesis, with a comma separating them, and they seem to think that
therefore something is well defined!

\textbf{Fan:} Even though causal inference has gained increasing attention
in statistics and beyond, there seems to be a lot of misunderstanding,
misuse, misinterpretation and mystifying of causal inference. Why? And what
needs to be done to change?

\textbf{Don:} I think it is partly because causal inference is a very
different topic from many topics in statistics in that it does not demand a
lot of technical advanced mathematical knowledge, but does demand a lot of
conceptual and basic mathematical sophistication. Principal Stratification
is one such example. Writing down notation does not take the place of
understanding what the notation means and how to prove things
mathematically. Also partly because causal inference has become a popular
topic, it has been flooded with publications that are often done casually.
For some fields, it is important to bridge the ``old''
(everything-based-on-OLS) thinking with the newer ideas. That's a battle
Guido and I constantly had to deal with when writing our book (Imbens and
Rubin, \citeyear{ImbRub14}).

\textbf{Fan:} You mentioned \textit{the} book; when will it finally come
out? It has been forthcoming for the last ten years or so.

\textbf{Don:} (Laughing) Come on, Fan, that's not fair! Has it only been
ten years? We have promised the publisher (Cambridge University Press) that
it will be ready by September 30, 2013.\footnote{As of April 1, 2014, the
book can be preordered on \href{http://www.amazon.com}{Amazon.com}.} It will be about 500 pages, 25
chapters. It will be followed by another volume, dealing with topics that
we could not get to in the volume due to length, such as principal
stratification beyond IV settings, or because we believe the topics have
not been sharply and cleanly formulated yet, such as regression
discontinuity designs, or using propensity scores with multiple treatments.
Also in this volume, we didn't discuss so-called case--control studies,
which are the meat of much of epidemiology; it is very important to embed
these studies into a framework that makes sense, not just teach them as a
bag of tricks.

\section*{Mentoring, Consulting and Editorship}

\textbf{Fabri:} You have advised over 50 Ph.D. students and many BA students
as well. This sounds like a job interview, but what is your teaching
philosophy?

\textbf{Don:} My view is that one should approach teaching very differently
depending on the kind of students you have and their goals. Harvard has
tremendous undergraduate and graduate students, but their strengths vary
and their objectives vary. A long time ago I decided that I don't have the
desire or ability to be an entertainer in class, that is, to entertain to
get their attention. If they find me entertaining, fine; but it is better
if they find the topic I am presenting entertaining.

\textbf{Fabri:} Many of your students went on to become leaders and not
only in academia. And you often say that the thing that you are the most
proud of is your students. Though it is clearly impossible to talk about
them here one by one, can you share some of your fond memories of the
students?

\textbf{Don:} Fabri, that is a killer question unless we have another day
for this. What I can say is that it has been a great pleasure to supervise
so  many very talented students. I could start listing my superb Ph.D. students at
the University of Chicago\vadjust{\goodbreak} and at Harvard. All of my Ph.D. students are
talented in many, and sometimes different, dimensions: among them there are
two COPSS award winners, one president of the ASA, one president of ENAR,
two JSM program chairs, and other such honors, and many of them made
substantial contributions to government, academia and industry.

\textbf{Fan:} You also have advised a large number of undergraduate
students on a wide range of topics. This is quite uncommon because some
people find mentoring undergraduates more challenging and less rewarding
than mentoring graduate students. What is your take on this?

\textbf{Don:} I am not completely innocent on this charge. I~have no
interest in ``babysitting'' and trying to motivate unmotivated students,
either undergraduate or graduate. But Harvard does attract some extremely
talented and motivated undergraduates, some of whom I had the pleasure to
advise. Five have won Hoopes and other prizes for outstanding undergraduate
theses.

\textbf{Fabri:} Now let's talk about writing, which both Fan and I, as many
others, have some quite memorable first-hand experience. You are known as a
perfectionist in writing. As you mentioned, you are willing to withdraw
accepted papers if you are not a hundred percent satisfied with them.

\textbf{Don:} Yes, as you guys know, I am generally a pain in the neck as a
coauthor. I have withdrawn three accepted papers, and tried to improve
them; all eventually got reaccepted. One of these is the paper with you
guys and others on multiple imputation for the CDC Anthrax vaccine trial
(Li et al., \citeyear{Lietal14}). You were not too happy about it initially.
%
\begin{figure*}

\includegraphics{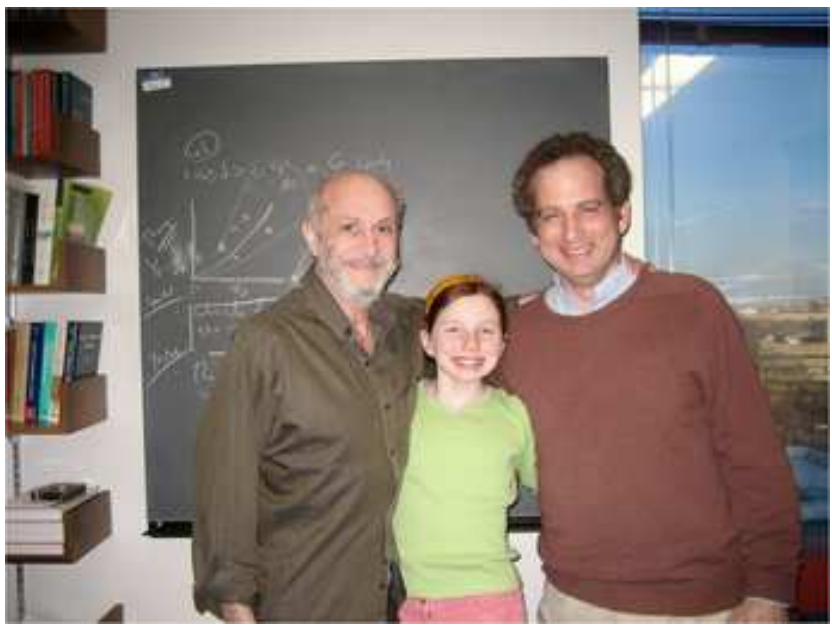}

\caption{D. B. Rubin (on left) with Tom Belin (on right) and Tom's daughter Janet
(middle), Cambridge, 2008.}
\end{figure*}

\textbf{Fabri:} (Laughing) Yeah, we tried to revolt without success. A
different question: How do you approach rejections? Do you have some advice
for young statisticians on that?

\textbf{Don:} Over the years I had many papers immediately rejected or
rejected with the suggestion that it would not be wise to resubmit.
However, in almost all of these cases, this treatment led to markedly
improved publications, somewhere. In fact, I think that the drafts that
have been repeatedly rejected possibly represent my best contributions.
Certainly, the repeated rejections, combined with my trying to address
various comments, led to better exposition and sometimes better problem
formulation, too. The most important idea is: Do not think that people who
are critics are hostile. In the vast majority of cases, editors and
reviewers are giving up their time\vadjust{\goodbreak} to try to \textit{help} authors, and, I
believe, are often especially generous and helpful to younger or
inexperienced authors. Do not read into rejection letters personal attacks,
which are extremely rare. So my advice is: Quality trumps quantity, and
stick with good ideas even when you have to do polite battle with editors
and reviewers---they are not perfect judges, but they are, almost
uniformly, on your side. More details of these are given in Rubin
(\citeyear{Rub14N2}).

\textbf{Fan:} In 1978, you became the Coordinating and Applications Editor
of JASA. Is there anything particularly unique about your editorship?

\textbf{Don:} As author, I am willing to withdraw accepted papers. As a new
editor, at least then, I was also willing to suggest to authors that they
withdraw papers accepted by the previous editors! I took some heat for that
at the beginning. I read through all the papers that the previous editorial
board had accepted and were awaiting copyediting for publication; for the
ones that I thought were bad (I remember there were about eight), I wrote,
``Dear authors, I think you should consider withdrawing this paper,'' with
long explanations of why I thought it would be an embarrassment to them if
the paper were published. Fabri knows that I can be brutally frank about
such suggestions.
%
\begin{figure*}

\includegraphics{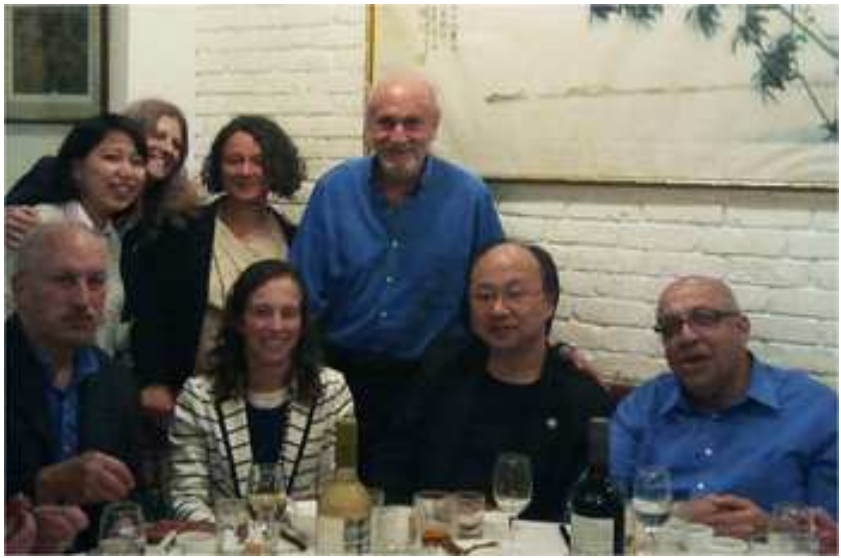}

\caption{Celebrating Don's 70th birthday at the Yenching
Restaurant, Harvard Square, March 29, 2014. Front (left to right): Alan
Zaslavsky, Elizabeth Stuart, Xiao-Li Meng, TE Raghunathan; Back (left to
right): Fan Li, Elizabeth Zell, Fabrizia Mealli, Don Rubin. The restaurant
has a dish named in Don's honor, the ``Rubin.''}
\end{figure*}

\textbf{Fan:} Did they comply?

\textbf{Don:} Yes, all but one. This one author fought, and I kept saying,
``You have to fix this up.'' Eventually, the changes made the paper OK. For
the other ones, the authors agreed with my criticisms:\vadjust{\goodbreak} Just because the
previous editor didn't get a good reviewer or they overlooked mistakes,
does not mean the paper should appear. But I was not very popular, at least
at first.

\textbf{Fabri:} You have done a wide range of consulting. What is the role
that consulting plays in your research?

\textbf{Don:} To me consulting is always a stimulating source of problems.
As I mentioned before, for example, propensity score technology partly came
from the consulting work we did for June Reinisch.

\textbf{Fabri:} One of the more controversial cases in which you are
involved as a consultant is the US tobacco litigation case, in which you
represented the tobacco\vadjust{\goodbreak} companies as an expert witness. Would you mind
sharing some of your thoughts on this case?

\textbf{Don:} Happy to. This comes from my family background dealing with
lawyers. We have a legal system where certain things are legal, certain
things are not. You should generally obey laws even if you don't like them,
or you should try to change them. If a company is making a legal product,
and they are advertising it legally under current laws, then accept it or
work to change the laws. If they lie, punish them for lying, if that is
legal to do. You never see a commercial for sporty cars that show the cars
going around corners extremely slowly and safely. How do they advertise
cars? They usually show them sweeping around corners, and say ``Don't do\vadjust{\goodbreak}
this on your own.'' Things that are enjoyable typically have uncertainties
or risks associated with them. Flying to Europe to visit Fabri has risks!

Certainly I do not doubt that no matter how I would intervene to reduce
cigarette smoking, lung cancer rates would drop. But what intervention that
would reduce smoking would involve reducing illegal conduct of the
cigarette industry---that is the essence of the legal question.

When I was first contacted by a tobacco lawyer, I~was very reluctant to
consult for them, and I feared strong pressure to be dishonest, which was
absent throughout. The original topic was simply to comment on the ways the
plaintiffs' experts were handling missing data. On examination, their
methods seemed to me to be not the best available and, at worst, silly
(e.g., when missing ``marital status,'' call them ``married''). As I
continued to read these initial reports, I was appalled that hundreds of
billions of dollars could be sought on the basis of such analyses. From a
broader perspective, the logic underlying most of the analyses also seemed
to me entirely confused. For example, alleged misconduct seemed to play no
role in nearly all calculations, and phrases such as ``caused by'' or
``attributable to,'' were used nearly interchangeably and often apparently
without thought. Should nearly a trillion dollars in damages be awarded on
the basis of faulty logic and bad statistical analyses because we ``know''
the defendant is evil and guilty? If the issue were assessing the tobacco
industry a trillion dollar fine for lying about its products, I would be
amazed but mute. But these reports were using statistical arguments to set
the numbers---is it acceptable to use bad statistics to set numbers
because we ``know'' the defendant is guilty? What sort of precedent does
that imply? The ethics of this consulting is discussed at some length in
Rubin (\citeyear{Rub02}).

\textbf{Fabri:} We have talked quite a lot about statistics. Let's talk
about some of your other passions in life, for example, music, audio
systems and sports cars.

\textbf{Don:} There are other passions, too, and their order is very age
dependent (I leave more to your perceptions). When a kid, for example,
sports cars, both driving them and rebuilding them, was the top of those
three hobbies. But age (poorer vision, slower reflexes, more aches and
pains, etc.) shifted the balance more to music, both live and recorded---luckily my ears are still good enough to enjoy these, but as more age
catches up, things may shift.

\textbf{Fan and Fabri:} Well, it has been nearly three hours since we
started the conversation. Here is the final question before letting you go
for dinner: What is your short advice to young researchers in statistics?

\textbf{Don:} Have fun! Don't be grumpy. If lucky, you may live to have a
wonderful 70th birthday celebration!\footnote{Video of the
celebration is available at: \url{http://www.stat.harvard.edu/DonRubin70/}}

\section*{Acknowledgments}
We thank Elizabeth Zell, Guido Imbens, Tom Belin, Rod Little, Dale Rinkel and Alan
Zaslavsky for helpful suggestions. This work is partially funded by NSF-SES
Grant 1155697.




\begin{thebibliography}{50}

\bibitem[\protect\citeauthoryear{Angrist, Imbens and Rubin}{1996}]{AngImbRub96}
\begin{barticle}[auto:STB|2014/06/18|12:29:53]
\bauthor{\bsnm{Angrist},~\bfnm{J.}\binits{J.}},
\bauthor{\bsnm{Imbens},~\bfnm{G.~W.}\binits{G.~W.}} \AND
\bauthor{\bsnm{Rubin},~\bfnm{D.~B.}\binits{D.~B.}}
(\byear{1996}).
\btitle{Identification of causal effects using instrumental variables (with discussion and rejoinder)}.
\bjournal{J. Amer. Statist. Assoc.}
\bvolume{91}
\bpages{444--472}.
\end{barticle}
\bptok{imsref}%
\endbibitem

\bibitem[\protect\citeauthoryear{Box}{1980}]{Box80}
\begin{barticle}[mr]
\bauthor{\bsnm{Box},~\bfnm{George~E.~P.}\binits{G.~E.~P.}}
(\byear{1980}).
\btitle{Sampling and {B}ayes' inference in scientific modelling and robustness}.
\bjournal{J. Roy. Statist. Soc. Ser. A}
\bvolume{143}
\bpages{383--430}.
\bid{doi={10.2307/2982063}, issn={0035-9238}, mr={0603745}}
\bptnote{check related}%
\end{barticle}
\bptok{imsref}%
\endbibitem

\bibitem[\protect\citeauthoryear{Brillinger, Jones and Tukey}{1978}]{Bri}
\begin{bincollection}[auto:STB|2014/06/18|12:29:53]
\bauthor{\bsnm{Brillinger},~\bfnm{D.~R.}\binits{D.~R.}},
\bauthor{\bsnm{Jones},~\bfnm{L.~V.}\binits{L.~V.}} \AND
\bauthor{\bsnm{Tukey},~\bfnm{J. W.}\binits{J. W.}}
(\byear{1978}).
\btitle{The management of weather resources}.
In \bbooktitle{The Role of Statistics in Weather Resources Management}
\bvolume{II}.
\bnote{Report of the Statistical Task Force to the Weather Modification Advisory
Board}.
\bpublisher{US Government Printing Office},
\blocation{Washington, DC}.
\end{bincollection}
\bptok{imsref}%
\endbibitem

\bibitem[\protect\citeauthoryear{Dempster, Laird and Rubin}{1977}]{DemLaiRub77}
\begin{barticle}[mr]
\bauthor{\bsnm{Dempster},~\bfnm{A.~P.}\binits{A.~P.}},
\bauthor{\bsnm{Laird},~\bfnm{N.~M.}\binits{N.~M.}} \AND
\bauthor{\bsnm{Rubin},~\bfnm{D.~B.}\binits{D.~B.}}
(\byear{1977}).
\btitle{Maximum likelihood from incomplete data via the EM algorithm}.
\bjournal{J.~R. Stat. Soc. Ser. B Stat. Methodol.}
\bvolume{39}
\bpages{1--38}.
\bid{issn={0035-9246}, mr={0501537}}
\bptnote{check related}%
\end{barticle}
\bptok{imsref}%
\endbibitem


\bibitem[\protect\citeauthoryear{Frangakis and Rubin}{2002}]{FraRub02}
\begin{barticle}[mr]
\bauthor{\bsnm{Frangakis},~\bfnm{Constantine~E.}\binits{C.~E.}} \AND
\bauthor{\bsnm{Rubin},~\bfnm{Donald~B.}\binits{D.~B.}}
(\byear{2002}).
\btitle{Principal stratification in causal inference}.
\bjournal{Biometrics}
\bvolume{58}
\bpages{21--29}.
\bid{doi={10.1111/j.0006-341X.2002.00021.x}, issn={0006-341X}, mr={1891039}}
\end{barticle}
\bptok{imsref}%
\endbibitem

\bibitem[\protect\citeauthoryear{Frumento et~al.}{2012}]{Fruetal12}
\begin{barticle}[mr]
\bauthor{\bsnm{Frumento},~\bfnm{Paolo}\binits{P.}},
\bauthor{\bsnm{Mealli},~\bfnm{Fabrizia}\binits{F.}},
\bauthor{\bsnm{Pacini},~\bfnm{Barbara}\binits{B.}} \AND
\bauthor{\bsnm{Rubin},~\bfnm{Donald~B.}\binits{D.~B.}}
(\byear{2012}).
\btitle{Evaluating the effect of training on wages in the presence of noncompliance, nonemployment, and missing outcome data}.
\bjournal{J. Amer. Statist. Assoc.}
\bvolume{107}
\bpages{450--466}.
\bid{doi={10.1080/01621459.2011.643719}, issn={0162-1459}, mr={2980057}}
\end{barticle}
\bptok{imsref}%
\endbibitem

\bibitem[\protect\citeauthoryear{Gelman and Rubin}{1992}]{GelRub92}
\begin{barticle}[auto:STB|2014/06/18|12:29:53]
\bauthor{\bsnm{Gelman},~\bfnm{A.}\binits{A.}} \AND
\bauthor{\bsnm{Rubin},~\bfnm{D.~B.}\binits{D.~B.}}
(\byear{1992}).
\btitle{Inference from iterative simulation using
multiple sequences (with discussion)}.
\bjournal{Statist. Sci.}
\bvolume{7}
\bpages{457--472}.
\end{barticle}
\bptok{imsref}%
\endbibitem


\bibitem[\protect\citeauthoryear{Gelman, Meng and Stern}{1996}]{G96}
\begin{barticle}[mr]
\bauthor{\bsnm{Gelman},~\bfnm{A.}\binits{A.}},
\bauthor{\bsnm{Meng},~\bfnm{X. L.}\binits{X. L.}} \AND
\bauthor{\bsnm{Stern},~\bfnm{H.}\binits{H.}}
(\byear{1996}).
\btitle{Posterior predictive assessment of model fitness via
realized discrepancies}.
\bjournal{Statist. Sinica}
\bvolume{6}
\bpages{733--760}.
\bid{mr={1422404}}
\end{barticle}
\bptok{imsref}%
\endbibitem



\bibitem[\protect\citeauthoryear{Gelman et~al.}{1995}]{Geletal95}
\begin{bbook}[mr]
\bauthor{\bsnm{Gelman},~\bfnm{Andrew}\binits{A.}},
\bauthor{\bsnm{Carlin},~\bfnm{John~B.}\binits{J.~B.}},
\bauthor{\bsnm{Stern},~\bfnm{Hal~S.}\binits{H.~S.}} \AND
\bauthor{\bsnm{Rubin},~\bfnm{Donald~B.}\binits{D.~B.}}
(\byear{1995}).
\btitle{Bayesian Data Analysis}.
\bpublisher{Chapman \& Hall},
\blocation{London}.
\bid{mr={1385925}}
\end{bbook}
\bptok{imsref}%
\endbibitem

\bibitem[\protect\citeauthoryear{Gelman et~al.}{2003}]{Geletal03}
\begin{bbook}[auto:STB|2014/06/18|12:29:53]
\bauthor{\bsnm{Gelman},~\bfnm{A.}\binits{A.}},
\bauthor{\bsnm{Carlin},~\bfnm{J.}\binits{J.}},
\bauthor{\bsnm{Stern},~\bfnm{H.}\binits{H.}} \AND
\bauthor{\bsnm{Rubin},~\bfnm{D.~B.}\binits{D.~B.}}
(\byear{2003}).
\btitle{Bayesian Data Analysis},
\bedition{2nd} ed.
\bpublisher{CRC Press},
\blocation{New York}.
\end{bbook}
\bptok{imsref}%
\endbibitem

\bibitem[\protect\citeauthoryear{Gelman et~al.}{2014}]{Geletal14}
\begin{bbook}[auto:STB|2014/06/18|12:29:53]
\bauthor{\bsnm{Gelman},~\bfnm{A.}\binits{A.}},
\bauthor{\bsnm{Carlin},~\bfnm{J.}\binits{J.}},
\bauthor{\bsnm{Stern},~\bfnm{H.}\binits{H.}},
\bauthor{\bsnm{Dunson},~\bfnm{D.}\binits{D.}},
\bauthor{\bsnm{Vehtari},~\bfnm{A.}\binits{A.}} \AND
\bauthor{\bsnm{Rubin},~\bfnm{D.~B.}\binits{D.~B.}}
(\byear{2014}).
\btitle{Bayesian Data Analysis},
\bedition{3rd} ed.
\bpublisher{CRC Press},
\blocation{New York}.
\end{bbook}
\bptok{imsref}%
\endbibitem

\bibitem[\protect\citeauthoryear{Glynn, Laird and Rubin}{1986}]{GlyLaiRub86}
\begin{bincollection}[auto:STB|2014/06/18|12:29:53]
\bauthor{\bsnm{Glynn},~\bfnm{R.}\binits{R.}},
\bauthor{\bsnm{Laird},~\bfnm{N.~M.}\binits{N.~M.}} \AND
\bauthor{\bsnm{Rubin},~\bfnm{D.~B.}\binits{D.~B.}}
(\byear{1986}).
\btitle{Selection modelling versus mixture modelling with nonignorable nonresponse}.
In \bbooktitle{Drawing Inferences from Self-Selected Samples}
(\beditor{\bfnm{H.}\binits{H.}~\bsnm{Wainer}}, ed.)
\bpages{119--146}.
\bpublisher{Springer},
\blocation{New York}.
\end{bincollection}
\bptok{imsref}%
\endbibitem

\bibitem[\protect\citeauthoryear{Hartley}{1956}]{Har56}
\begin{barticle}[mr]
\bauthor{\bsnm{Hartley},~\bfnm{H.~O.}\binits{H.~O.}}
(\byear{1956}).
\btitle{A plan for programming analysis of variance for general purpose computers}.
\bjournal{Biometrics}
\bvolume{12}
\bpages{110--122}.
\bid{issn={0006-341X}, mr={0079359}}
\end{barticle}
\bptok{imsref}%
\endbibitem

\bibitem[\protect\citeauthoryear{Hartley and Hocking}{1971}]{HarHoc71}
\begin{barticle}[auto:STB|2014/06/18|12:29:53]
\bauthor{\bsnm{Hartley},~\bfnm{H.~O.}\binits{H.~O.}} \AND
\bauthor{\bsnm{Hocking},~\bfnm{R.~R.}\binits{R.~R.}}
(\byear{1971}).
\btitle{The analysis of incomplete data}.
\bjournal{Biometrics}
\bvolume{27}
\bpages{783--823}.
\end{barticle}
\bptok{imsref}%
\endbibitem

\bibitem[\protect\citeauthoryear{Holland}{1986}]{Hol86}
\begin{barticle}[mr]
\bauthor{\bsnm{Holland},~\bfnm{Paul~W.}\binits{P.~W.}}
(\byear{1986}).
\btitle{Statistics and causal inference}.
\bjournal{J. Amer. Statist. Assoc.}
\bvolume{81}
\bpages{945--970}.
\bid{issn={0162-1459}, mr={0867618}}
\bptnote{check related}%
\end{barticle}
\bptok{imsref}%
\endbibitem

\bibitem[\protect\citeauthoryear{Imbens and Angrist}{1994}]{ImbAng94}
\begin{barticle}[auto:STB|2014/06/18|12:29:53]
\bauthor{\bsnm{Imbens},~\bfnm{G.~W.}\binits{G.~W.}} \AND
\bauthor{\bsnm{Angrist},~\bfnm{J.}\binits{J.}}
(\byear{1994}).
\btitle{Identification and estimation of local average treatment effects}.
\bjournal{Econometrica}
\bvolume{62}
\bpages{467--475}.
\end{barticle}
\bptok{imsref}%
\endbibitem

\bibitem[\protect\citeauthoryear{Imbens and Rubin}{1997}]{ImbRub97}
\begin{barticle}[mr]
\bauthor{\bsnm{Imbens},~\bfnm{Guido~W.}\binits{G.~W.}} \AND
\bauthor{\bsnm{Rubin},~\bfnm{Donald~B.}\binits{D.~B.}}
(\byear{1997}).
\btitle{Bayesian inference for causal effects in randomized experiments with noncompliance}.
\bjournal{Ann. Statist.}
\bvolume{25}
\bpages{305--327}.
\bid{doi={10.1214/aos/1034276631}, issn={0090-5364}, mr={1429927}}
\end{barticle}
\bptok{imsref}%
\endbibitem

\bibitem[\protect\citeauthoryear{Imbens and Rubin}{2015}]{ImbRub14}
\begin{bbook}[auto:STB|2014/06/18|12:29:53]
\bauthor{\bsnm{Imbens},~\bfnm{G.~W.}\binits{G.~W.}} \AND
\bauthor{\bsnm{Rubin},~\bfnm{D.~B.}\binits{D.~B.}}
(\byear{2015}).
\btitle{Causal Inference for Statistics, Social, and Biomedical Sciences: An Introduction}.
\bpublisher{Cambridge Univ. Press},
\blocation{New York}.
\end{bbook}
\bptok{imsref}%
\endbibitem

\bibitem[\protect\citeauthoryear{Li et~al.}{2014}]{Lietal14}
\begin{barticle}[mr]
\bauthor{\bsnm{Li},~\bfnm{Fan}\binits{F.}},
\bauthor{\bsnm{Baccini},~\bfnm{Michela}\binits{M.}},
\bauthor{\bsnm{Mealli},~\bfnm{Fabrizia}\binits{F.}},
\bauthor{\bsnm{Zell},~\bfnm{Elizabeth~R.}\binits{E.~R.}},
\bauthor{\bsnm{Frangakis},~\bfnm{Constantine~E.}\binits{C.~E.}} \AND
\bauthor{\bsnm{Rubin},~\bfnm{Donald~B.}\binits{D.~B.}}
(\byear{2014}).
\btitle{Multiple {i}mputation by {o}rdered {m}onotone {b}locks {w}ith {a}pplication to the {a}nthrax {v}accine {r}esearch {p}rogram}.
\bjournal{J. Comput. Graph. Statist.}
\bvolume{23}
\bpages{877--892}.
\bid{doi={10.1080/10618600.2013.826583}, issn={1061-8600}, mr={3224660}}
\end{barticle}
\bptok{imsref}%
\endbibitem

\bibitem[\protect\citeauthoryear{Little and Rubin}{1987}]{LitRub87}
\begin{bbook}[mr]
\bauthor{\bsnm{Little},~\bfnm{Roderick~J.~A.}\binits{R.~J.~A.}} \AND
\bauthor{\bsnm{Rubin},~\bfnm{Donald~B.}\binits{D.~B.}}
(\byear{1987}).
\btitle{Statistical Analysis with Missing Data}.
\bpublisher{Wiley},
\blocation{New York}.
\bid{mr={0890519}}
\end{bbook}
\bptok{imsref}%
\endbibitem

\bibitem[\protect\citeauthoryear{Little and Rubin}{2002}]{LitRub02}
\begin{bbook}[mr]
\bauthor{\bsnm{Little},~\bfnm{Roderick~J.~A.}\binits{R.~J.~A.}} \AND
\bauthor{\bsnm{Rubin},~\bfnm{Donald~B.}\binits{D.~B.}}
(\byear{2002}).
\btitle{Statistical Analysis with Missing Data},
\bedition{2nd} ed.
\bpublisher{Wiley},
\blocation{Hoboken, NJ}.
\bid{mr={1925014}}
\end{bbook}
\bptok{imsref}%
\endbibitem

\bibitem[\protect\citeauthoryear{Mattei, Li and Mealli}{2013}]{MatLiMea13}
\begin{barticle}[mr]
\bauthor{\bsnm{Mattei},~\bfnm{Alessandra}\binits{A.}},
\bauthor{\bsnm{Li},~\bfnm{Fan}\binits{F.}} \AND
\bauthor{\bsnm{Mealli},~\bfnm{Fabrizia}\binits{F.}}
(\byear{2013}).
\btitle{Exploiting multiple outcomes in {B}ayesian principal stratification analysis with application to the evaluation of a job training program}.
\bjournal{Ann. Appl. Stat.}
\bvolume{7}
\bpages{2336--2360}.
\bid{doi={10.1214/13-AOAS674}, issn={1932-6157}, mr={3161725}}
\end{barticle}
\bptok{imsref}%
\endbibitem

\bibitem[\protect\citeauthoryear{Mealli and Pacini}{2013}]{MeaPac13}
\begin{barticle}[mr]
\bauthor{\bsnm{Mealli},~\bfnm{Fabrizia}\binits{F.}} \AND
\bauthor{\bsnm{Pacini},~\bfnm{Barbara}\binits{B.}}
(\byear{2013}).
\btitle{Using {s}econdary {o}utcomes to {s}harpen {i}nference in {r}andomized {e}xperiments {w}ith {n}oncompliance}.
\bjournal{J. Amer. Statist. Assoc.}
\bvolume{108}
\bpages{1120--1131}.
\bid{doi={10.1080/01621459.2013.802238}, issn={0162-1459}, mr={3174688}}
\end{barticle}
\bptok{imsref}%
\endbibitem

\bibitem[\protect\citeauthoryear{Meng}{1994}]{Men94}
\begin{barticle}[mr]
\bauthor{\bsnm{Meng},~\bfnm{Xiao-Li}\binits{X.-L.}}
(\byear{1994}).
\btitle{Posterior predictive {$p$}-values}.
\bjournal{Ann. Statist.}
\bvolume{22}
\bpages{1142--1160}.
\bid{doi={10.1214/aos/1176325622}, issn={0090-5364}, mr={1311969}}
\end{barticle}
\bptok{imsref}%
\endbibitem

\bibitem[\protect\citeauthoryear{Neyman}{1990}]{Spl90}
\begin{barticle}[mr]
\bauthor{\bsnm{Neyman},~\bfnm{Jerzy}\binits{J.}}
(\byear{1990}).
\btitle{On the application of probability theory to agricultural experiments. {E}ssay on principles. {S}ection~9}.
\bjournal{Statist. Sci.}
\bvolume{5}
\bpages{465--472}.
\bnote{Translated from the Polish and edited by D. M. D\c{a}browska and T. P. Speed}.
\bid{issn={0883-4237}, mr={1092986}}
\end{barticle}
\bptok{imsref}%
\endbibitem

\bibitem[\protect\citeauthoryear{Neyman}{1934}]{Ney34}
\begin{barticle}[auto:STB|2014/06/18|12:29:53]
\bauthor{\bsnm{Neyman},~\bfnm{J.}\binits{J.}}
(\byear{1934}).
\btitle{On the two different aspects of the representative method: The method of stratified sampling and the method of purposive selection}.
\bjournal{J. Roy. Statist. Soc.}
\bvolume{97}
\bpages{558--625}.
\end{barticle}
\bptok{imsref}%
\endbibitem

\bibitem[\protect\citeauthoryear{Rosenbaum and Rubin}{1983a}]{RosRub83}
\begin{barticle}[mr]
\bauthor{\bsnm{Rosenbaum},~\bfnm{Paul~R.}\binits{P.~R.}} \AND
\bauthor{\bsnm{Rubin},~\bfnm{Donald~B.}\binits{D.~B.}}
(\byear{1983a}).
\btitle{The central role of the propensity score in observational studies for causal effects}.
\bjournal{Biometrika}
\bvolume{70}
\bpages{41--55}.
\bid{doi={10.1093/biomet/70.1.41}, issn={0006-3444}, mr={0742974}}
\end{barticle}
\bptok{imsref}%
\endbibitem

\bibitem[\protect\citeauthoryear{Rosenbaum and Rubin}{1983b}]{autokey27}
\begin{barticle}[auto:STB|2014/06/18|12:29:53]
\bauthor{\bsnm{Rosenbaum},~\bfnm{Paul~R.}\binits{P.~R.}} \AND
\bauthor{\bsnm{Rubin},~\bfnm{Donald~B.}\binits{D.~B.}}
(\byear{1983b}).
\btitle{Assessing sensitivity to an unobserved binary covariate in an observational study with binary outcome}.
\bjournal{J. R. Stat. Soc. Ser. B Stat. Methodol.}
\bvolume{45}
\bpages{212--218}.
\end{barticle}
\bptok{imsref}%
\endbibitem

\bibitem[\protect\citeauthoryear{Rosenbaum and Rubin}{1984a}]{autokey28}
\begin{barticle}[auto:STB|2014/06/18|12:29:53]
\bauthor{\bsnm{Rosenbaum},~\bfnm{Paul~R.}\binits{P.~R.}} \AND
\bauthor{\bsnm{Rubin},~\bfnm{Donald~B.}\binits{D.~B.}}
(\byear{1984a}).
\btitle{Reducing bias in observational studies using subclassification on the propensity score}.
\bjournal{J. Amer. Statist. Assoc.}
\bvolume{79}
\bpages{516--524}.
\end{barticle}
\bptok{imsref}%
\endbibitem

\bibitem[\protect\citeauthoryear{Rosenbaum and Rubin}{1984b}]{autokey29}
\begin{barticle}[auto:STB|2014/06/18|12:29:53]
\bauthor{\bsnm{Rosenbaum},~\bfnm{Paul~R.}\binits{P.~R.}} \AND
\bauthor{\bsnm{Rubin},~\bfnm{Donald~B.}\binits{D.~B.}}
(\byear{1984b}).
\btitle{Sensitivity of Bayes inference with data-dependent stopping rules}.
\bjournal{Amer. Statist.}
\bvolume{38}
\bpages{106--109}.
\end{barticle}
\bptok{imsref}%
\endbibitem

\bibitem[\protect\citeauthoryear{Rubin}{1972}]{Rub72}
\begin{barticle}[mr]
\bauthor{\bsnm{Rubin},~\bfnm{Donald~B.}\binits{D.~B.}}
(\byear{1972}).
\btitle{A non-iterative algorithm for least squares estimation of missing values in any analysis of variance design}.
\bjournal{J. R. Stat. Soc. Ser. C. Appl. Stat.}
\bvolume{21}
\bpages{136--141}.
\bid{issn={0035-9254}, mr={0311040}}
\end{barticle}
\bptok{imsref}%
\endbibitem

\bibitem[\protect\citeauthoryear{Rubin}{1974}]{Rub74}
\begin{barticle}[auto:STB|2014/06/18|12:29:53]
\bauthor{\bsnm{Rubin},~\bfnm{D.~B.}\binits{D.~B.}}
(\byear{1974}).
\btitle{Estimating causal effects of treatments in randomized and nonrandomized studies}.
\bjournal{J. Educational Psychology}
\bvolume{66}
\bpages{688--701}.
\end{barticle}
\bptok{imsref}%
\endbibitem

\bibitem[\protect\citeauthoryear{Rubin}{1976}]{Rub76}
\begin{barticle}[mr]
\bauthor{\bsnm{Rubin},~\bfnm{Donald~B.}\binits{D.~B.}}
(\byear{1976}).
\btitle{Inference and missing data}.
\bjournal{Biometrika}
\bvolume{63}
\bpages{581--592}.
\bid{issn={0006-3444}, mr={0455196}}
\bptnote{check related}%
\end{barticle}
\bptok{imsref}%
\endbibitem

\bibitem[\protect\citeauthoryear{Rubin}{1977}]{Rub77}
\begin{barticle}[auto:STB|2014/06/18|12:29:53]
\bauthor{\bsnm{Rubin},~\bfnm{D.~B.}\binits{D.~B.}}
(\byear{1977}).
\btitle{Assignment to treatment group on the basis of a covariate}.
\bjournal{J. Educational Statistics}
\bvolume{2}
\bpages{1--26}.
\end{barticle}
\bptok{imsref}%
\endbibitem

\bibitem[\protect\citeauthoryear{Rubin}{1978a}]{Rub78N1}
\begin{barticle}[mr]
\bauthor{\bsnm{Rubin},~\bfnm{Donald~B.}\binits{D.~B.}}
(\byear{1978a}).
\btitle{Bayesian inference for causal effects: The role of randomization}.
\bjournal{Ann. Statist.}
\bvolume{6}
\bpages{34--58}.
\bid{issn={0090-5364}, mr={0472152}}
\end{barticle}
\bptok{imsref}%
\endbibitem

\bibitem[\protect\citeauthoryear{Rubin}{1978b}]{Rub78N2}
\begin{bincollection}[auto:STB|2014/06/18|12:29:53]
\bauthor{\bsnm{Rubin},~\bfnm{D.~B.}\binits{D.~B.}}
(\byear{1978b}).
\btitle{Multiple imputations in sample surveys---A phenomenological Bayesian approach to nonresponse (with discussion and reply)}.
In \bbooktitle{The Proceedings of the Survey Research Methods Section of the American Statistical Association}
\bpages{20--34}.
\bnote{Also in \emph{Imputation and Editing of Faulty or Missing Survey Data}}.
\bpublisher{U.S. Dept. Commerce},
\blocation{Bureau of the Census, Washington, DC}.
\end{bincollection}
\bptok{imsref}%
\endbibitem

\bibitem[\protect\citeauthoryear{Rubin}{1981}]{Rub81}
\begin{barticle}[mr]
\bauthor{\bsnm{Rubin},~\bfnm{Donald~B.}\binits{D.~B.}}
(\byear{1981}).
\btitle{The {B}ayesian bootstrap}.
\bjournal{Ann. Statist.}
\bvolume{9}
\bpages{130--134}.
\bid{issn={0090-5364}, mr={0600538}}
\end{barticle}
\bptok{imsref}%
\endbibitem

\bibitem[\protect\citeauthoryear{Rubin}{1983}]{Rub83}
\begin{bincollection}[mr]
\bauthor{\bsnm{Rubin},~\bfnm{Donald~B.}\binits{D.~B.}}
(\byear{1983}).
\btitle{A case study of the robustness of {B}ayesian methods of inference: Estimating the total in a finite population using transformations to normality}.
In \bbooktitle{Scientific Inference, Data Analysis, and Robustness ({M}adison, {W}is., 1981)}.
\bseries{Publ. Math. Res. Center Univ. Wisconsin}
\bvolume{48}
\bpages{213--244}.
\bpublisher{Academic Press},
\blocation{Orlando, FL}.
\bid{mr={0772771}}
\end{bincollection}
\bptok{imsref}%
\endbibitem

\bibitem[\protect\citeauthoryear{Rubin}{1984}]{Rub84}
\begin{barticle}[mr]
\bauthor{\bsnm{Rubin},~\bfnm{Donald~B.}\binits{D.~B.}}
(\byear{1984}).
\btitle{Bayesianly justifiable and relevant frequency calculations for the applied statistician}.
\bjournal{Ann. Statist.}
\bvolume{12}
\bpages{1151--1172}.
\bid{doi={10.1214/aos/1176346785}, issn={0090-5364}, mr={0760681}}
\end{barticle}
\bptok{imsref}%
\endbibitem


\bibitem[\protect\citeauthoryear{Rubin}{1987a}]{Rub87N1}
\begin{bbook}[auto:STB|2014/06/18|12:29:53]
\bauthor{\bsnm{Rubin},~\bfnm{D.~B.}\binits{D.~B.}}
(\byear{1987a}).
\btitle{Multiple Imputation for Nonresponse in Surveys}.
\bpublisher{Wiley},
\blocation{New York}.
\end{bbook}
\bptok{imsref}%
\endbibitem

\bibitem[\protect\citeauthoryear{Rubin}{1987b}]{Rub87N2}
\begin{barticle}[auto:STB|2014/06/18|12:29:53]
\bauthor{\bsnm{Rubin},~\bfnm{D.~B.}\binits{D.~B.}}
(\byear{1987b}).
\btitle{A noniterative sampling/importance resampling alternative to the data augmentation algorithm for creating a few imputations when fractions of missing information are modest:
The SIR algorithm.\vadjust{\eject} Discussion of ``The calculation of posterior distributions by data augmentation'' by M. Tanner and W. H. Wong}.
\bjournal{J. Amer. Statist. Assoc.}
\bvolume{82}
\bpages{543--546}.
\end{barticle}
\bptok{imsref}%
\endbibitem

\bibitem[\protect\citeauthoryear{Rubin}{1990a}]{Rub90N1}
\begin{barticle}[auto:STB|2014/06/18|12:29:53]
\bauthor{\bsnm{Rubin},~\bfnm{D.~B.}\binits{D.~B.}}
(\byear{1990a}).
\btitle{Formal modes of statistical inference for causal effects}.
\bjournal{J. Statist. Plann. Inference}
\bvolume{25}
\bpages{279--292}.
\end{barticle}
\bptok{imsref}%
\endbibitem

\bibitem[\protect\citeauthoryear{Rubin}{1990b}]{Rub90N2}
\begin{barticle}[auto:STB|2014/06/18|12:29:53]
\bauthor{\bsnm{Rubin},~\bfnm{D.~B.}\binits{D.~B.}}
(\byear{1990b}).
\btitle{Comment on ``Neyman (1923) and causal inference in experiments and observational studies.''}
\bjournal{Statist. Sci.}
\bvolume{5}
\bpages{472--480}.
\bid{mr={1092987}}
\end{barticle}
\bptok{imsref}%
\endbibitem

\bibitem[\protect\citeauthoryear{Rubin}{1994}]{Efr94}
\begin{barticle}[auto:STB|2014/06/18|12:29:53]
\bauthor{\bsnm{Rubin},~\bfnm{D.~B.}\binits{D.~B.}}
(\byear{1994}).
\btitle{Comment on ``Missing data, imputation, and the bootstrap''
by Bradley Efron}.
\bjournal{J. Amer. Statist. Assoc.}
\bvolume{89}
\bpages{475--478}.
\end{barticle}
\bptok{imsref}%
\endbibitem



\bibitem[\protect\citeauthoryear{Rubin}{1995}]{Rub95}
\begin{barticle}[auto:STB|2014/06/18|12:29:53]
\bauthor{\bsnm{Rubin},~\bfnm{D.~B.}\binits{D.~B.}}
(\byear{1995}).
\btitle{Bayes, Neyman, and calibration. Discussion of Berk, Western and Weiss}.
\bjournal{Sociological Methodology}
\bvolume{25}
\bpages{473--479}.
\end{barticle}
\bptok{imsref}%
\endbibitem

\bibitem[\protect\citeauthoryear{Rubin}{1996}]{Rub96}
\begin{barticle}[auto:STB|2014/06/18|12:29:53]
\bauthor{\bsnm{Rubin},~\bfnm{D.~B.}\binits{D.~B.}}
(\byear{1996}).
\btitle{Multiple imputation after $18+$ years (with discussion and rejoinder)}.
\bjournal{J. Amer. Statist. Assoc.}
\bvolume{91}
\bpages{473--517}.
\end{barticle}
\bptok{imsref}%
\endbibitem

\bibitem[\protect\citeauthoryear{Rubin}{2002}]{Rub02}
\begin{barticle}[auto:STB|2014/06/18|12:29:53]
\bauthor{\bsnm{Rubin},~\bfnm{D.~B.}\binits{D.~B.}}
(\byear{2002}).
\btitle{The ethics of consulting for the tobacco industry. Special issue on ``Ethics, statistics and statisticians''}.
\bjournal{Stat. Methods Med. Res.}
\bvolume{11}
\bpages{373--380}.
\end{barticle}
\bptok{imsref}%
\endbibitem

\bibitem[\protect\citeauthoryear{Rubin}{2010}]{Rub10}
\begin{barticle}[pbm]
\bauthor{\bsnm{Rubin},~\bfnm{Donald~B.}\binits{D.~B.}}
(\byear{2010}).
\btitle{Reflections stimulated by the comments of Shadish (2010) and West and Thoemmes (2010)}.
\bjournal{Psychol. Methods}
\bvolume{15}
\bpages{38--46}.
\bid{doi={10.1037/a0018537}, issn={1939-1463}, mid={NIHMS184768}, pii={2010-04293-004}, pmcid={2891035}, pmid={20230101}}
\end{barticle}
\bptok{imsref}%
\endbibitem

\bibitem[\protect\citeauthoryear{Rubin}{2014a}]{Rub14N1}
\begin{bincollection}[auto:STB|2014/06/18|12:29:53]
\bauthor{\bsnm{Rubin},~\bfnm{D.~B.}\binits{D.~B.}}
(\byear{2014a}).
\btitle{Converting rejections into positive stimuli}.
In \bbooktitle{Past, Present, and Future of Statistical Science}
(\beditor{\binits{X.} \bsnm{Lin et al.}}, eds.)
\bpages{593--603}.
\bpublisher{CRC Press},
\blocation{New York}.
\end{bincollection}
\bptok{imsref}%
\endbibitem

\bibitem[\protect\citeauthoryear{Rubin}{2014b}]{Rub14N2}
\begin{bincollection}[auto:STB|2014/06/18|12:29:53]
\bauthor{\bsnm{Rubin},~\bfnm{D.~B.}\binits{D.~B.}}
(\byear{2014b}).
\btitle{The importance of mentors}.
In \bbooktitle{Past, Present, and Future of Statistical Science}
(\beditor{\binits{X.} \bsnm{Lin et al.}}, eds.)
\bpages{605--613}.
\bpublisher{CRC Press},
\blocation{New York}.
\end{bincollection}
\bptok{imsref}%
\endbibitem\vfill



\end{thebibliography}
\end{document}